\input jnl.tex
\input reforder.tex
\input epsf.tex

\def\a{{\alpha}}

\def\e{{\epsilon}}

\def\om{{\omega}}
\def\half{{1 \over 2}}
\def\ra{{\rangle}}
\def\la{{\langle}}

\def\l{\ell}

\def\p{{\bf p}}

\def\k{{\bf k}}

\def\x{{\bf x}}
\def\y{{\bf y}}

\def\ria{{\rightarrow}}

\def\l{{\lambda}}

\def\la{\langle}
\def\ra{\rangle}
\def\ria{\rightarrow}
\def\pp{\prime\prime}
\def\x{{\bf x}}
\def\y{{\bf y}}

\def\X'{{x^{\mu}}^{\prime}}

\def\R{{I\kern-0.3em R}}

\def\ula#1{\raise2ex\hbox{$\leftarrow$}\mkern-16.5mu #1}
%upleftarrow
\def\ura#1{\raise2ex\hbox{$\rightarrow$}\mkern-16.5mu #1}
%uprightarrow
\def\ulra#1{\raise1.5ex\hbox{$\leftrightarrow$}\mkern-16.5mu #1}
%upleftrightarrow
%\def\uplrarrow#1{#1^{{}^{\!\!\!\!\!\leftrightarrow}}}
%\def\lleftarrow#1{#1^{{}^{\!\!\!\!\!\!\leftarrow}}}
%\oneandahalfspace

\centerline{\bf Decoherent Histories Analysis of the Relativistic
Particle}

\vskip 0.2in
\author J.J.Halliwell % \footnote{$^{\dag}$}{E-mail address: \jjh}
\vskip 0.4in
\centerline{\rm and}
\author J.Thorwart

\vskip 0.2in
\affil Theory Group, Blackett Laboratory,
Imperial College, London SW7 2BZ, UK.

\vskip 0.5in
\centerline {\rm Preprint IC/TP/0-01/20. June, 2001}
%\vskip 0.1in
%\vskip 1.0in
%\centerline {\rm  PACS numbers: 03.65.-w, 03.65.Bz, 06.30.Ft}
%\vskip 1.0in

\abstract{The Klein-Gordon equation is a useful test arena for
quantum cosmological models described by the Wheeler-DeWitt
equation. We use the decoherent histories approach to quantum
theory to obtain the probability that a free relativistic particle
crosses a section of spacelike surface. The decoherence functional
is constructed using path integral methods with initial states
attached using the (positive definite) ``induced'' inner product
between solutions to the constraint equation. The construction is
complicated by the fact that the amplitudes (class operators)
calculated using a path integral typically do not satisfy the
constraint equation everywhere, but we show how they may be
systematically modified in such a way that they do satisfy the
constraint. The notion of crossing a spacelike surface requires
some attention, given that the paths in the path integral may
cross such a surface many times, but we show that first and last
crossings are in essence the only useful possibilities. Different
possible results for the probabilities are obtained, depending on
how the relativistic particle is quantized (using the Klein-Gordon
equation, or its square root, with the associated Newton-Wigner
states). In the Klein-Gordon quantization, the decoherence is only
approximate, due to the fact that the paths in the path integral
may go backwards and forwards in time. We compare with the results
obtained using operators which commute with the constraint (the
``evolving constants'' method).}

\endtopmatter
\endpage

\head{\bf 1. Introduction}

If the state of a quantum
system obeys a wave equation of the form
$$
H \Psi = 0,
\eqno(1.1)
$$
how do we extract probabilities from the wave function? Quantum
cosmological models are described by precisely such a wave
equation -- the Wheeler-DeWitt equation -- where $H$ is the total
Hamiltonian of the matter and gravitational fields
[\cite{Har3,HaHa}]. A simpler
but important example of an equation of this type is
the Klein-Gordon equation
$$
\left( \square + m^2 \right) \phi = 0
\eqno(1.2)
$$
The Wheeler-DeWitt equation for simple models has the form of
a Klein-Gordon equation in a general curved spacetime background
with a spacetime dependent mass term. Traditional approaches
to relativistic quantum theory note the various difficulties
of interpreting the Klein-Gordon equation and then pass
quickly on to quantum field theory. The Wheeler-DeWitt equation,
however, in its full form, already represents a second-quantized
field theory. To work with the Wheeler-DeWitt equation we must
therefore return to wave equations of the Klein-Gordon type
and understand how to overcome their difficulties without
resorting to second quantization. These questions are closely
related to the problem of time in quantum gravity [\cite{BuIs,Ish,Kuc}].
Furthermore, it is likely that they will also be present in
other approaches to quantum gravity, such as the loop variable
approach [\cite{Ash}].

A number of approaches to this problem have been considered. One
is the decoherent histories approach, in which the wave function
is associated with a set of histories to which probabilities may
be assigned [\cite{GeH,Gri,Omn}].
Another approach involves the use of operators which
commute with $H$ [\cite{Rov1,Rov2,Rov3,DeW,Mar1,Mar2}].
The purpose of this paper is to explore the
application of these methods to some simple probability questions
involving the relativistic particle in flat spacetime. We will
concentrate on the decoherent histories approach, extending and
developing earlier work in this area [\cite{Har3,HaMa}]. At the present (rather
early) stage of development of this field it is not possible to
say whether it is equivalent to an operator approach, and an
important part of this paper is to compare the approaches where
possible. Indeed, there are many different and potentially
inequivalent quantization methods applicable to simple
parametrized systems [\cite{AnSa,Hal2}].

We will focus on the following question: given a solution to the
Klein-Gordon, what is the probability of finding the particle in a
spatial region $\Delta$ of a spacelike surface? The question is
clearly a simple one, but it turns out to expose some subtle
aspects of the decoherent histories approach applied to
parameterized systems, and is an important test of the formalism
in a familiar situation. Furthermore, it may also be regarded as a
preparatory exercise for the treatment of more complicated quantum
cosmological models, which will be considered elsewhere.

We begin by briefly reviewing the various aspects of the formalism
relating to the wave equations (1.1), (1.2).

\subhead{\bf 1(A). Inner Products}

The inner product traditionally associated with the Klein--Gordon
and similar equations is the Klein-Gordon inner product,
$$
\eqalignno{
\Psi^* \circ_{KG} \Phi &= (\Psi, \Phi)_{KG} = i \int_{\Sigma} d^3x
\ \Psi^* \ulra{\partial}_0
\Phi
\cr &= i \int_{\Sigma} d^3 x \ \left( \Psi^* \ura{\partial}_0 \Phi - \Psi^*
\ula{\partial}_0 \Phi \right)
&(1.3) \cr}
$$
It is evaluated on a spacelike surface $\Sigma$, and
is independent of the choice of such surface if $\Psi$ and
$\Phi$ are solutions to the Klein-Gordon equation.
This inner product is, however, not positive definite. When a separation into
positive and negative frequencies is possible, it is positive
on the positive frequency sector and negative on the negative
frequency sector.

An essential part of our approach here is to use the
methods of refined algebraic quantization, in which we
work with the so-called induced
(or Rieffel) inner product [\cite{HaMa,Rie}]. This starts
through the introduction of an auxiliary inner product
$$
(\Psi, \Phi)_A = \int d^4x \ \Psi^* (x) \Phi (x)
\eqno(1.4)
$$
We then consider eigenstates of the constraint
$$
H \Psi_{\l k} = \l \Psi_{\l k}
\eqno(1.5)
$$
where $k$ is a degeneracy label. These are normalizable
in the auxiliary inner product via
$$
( \Psi_{\l k}, \Psi_{\l' k'} )_A =
\delta (\l - \l') \delta (k - k')
\eqno(1.6)
$$
The induced inner product between solutions to the constraint
then consists of dropping the $\delta (\l - \l')$ term on the
right and taking the limit $\l, \l' \ria 0 $. This produces a
well-defined positive definite inner product on solutions to the
constraint. This procedure may also be understood by starting from
the observation that solutions to the constraint (1.5) may be
written in the form $ \delta (H - \lambda ) | \chi \ra $ for some
fiducial state $ | \chi \ra $. The induced inner product between
such states is then effectively equivalent to replacing $ [
\delta (H -\lambda ) ]^2 $ with $ \delta (H - \lambda ) $,
that is, taking
$ \delta (H - \lambda) $ to be a projection operator (which it clearly
is in the case of a discrete spectrum).

When a split into positive and negative frequency solutions is
possible, $ \Psi = \Psi^+ + \Psi^- $, the induced inner product
coincides with the Klein-Gordon inner product but with the sign of the
negative frequency sector changed so as to make the product
positive:
$$
\Psi^* \circ_I \Phi = (\Psi, \Phi)_I = (\Psi^+, \Phi^+)_{KG}
- (\Psi^-, \Phi^-)_{KG}
\eqno(1.7)
$$
For the free relativistic particle, we could quite simply have
defined an inner product by the object on the right, and this is
well-defined since the positive and negative frequency sectors do
not interact in this case. The advantage of the induced inner
product, however, is that provides a good inner product even when
the split into positive and negative frequencies is not possible.
(See Refs.[\cite{HaMa,Rie}] for more details).

\subhead{\bf 1(B). An Operator Approach}

Given the constraint equation (1.1) and its inner product structure
we would like to be able to assign probabilities to various
dynamical variables of interest. It is generally believed that the
interesting dynamical variables are those that commute
with the constraint $H$
[\cite{Rov1,Rov2,Rov3,DeW,Mar1,Mar2}]. This is because the constraint
is associated with reparametrization invariance
(diffeomorphism invariance in the general case)
and we are interested in variables that are invariant.

Because the wave equation is not of the Schr\"odinger type,
it does not have an external time variable, so we cannot
talk about the value of a variable at a particular ``time''.
Instead, ``time'' is somehow encoded in the variables
already present in the wave equation. In the Klein-Gordon
equation, for example, we might be interested in
the value of the spatial coordinate $\x$ at given $x^0$.
We might equally be interested in
the value of $x^0$, say, at given $x^1$. We need
operators commuting with $H$ which express these
quantities.

Suppose we are interested in the operator corresponding
to the value of $A$ when $B$ takes the value $\tau$.
The appropriate operator is
$$
[A]_{B=\tau} = \int_{-\infty}^{\infty} ds \ A(s)
\ { d B (s) \over ds } \ \delta\left(
B(s) - \tau \right)
\eqno(1.8)
$$
where $A(s) = e^{ i Hs} A e^{- i H s } $, and similarly for $B(s)$
[\cite{Mar1,Mar2}]. (We assume a suitable operator ordering is
chosen in this expression, although note that it is not always
possible to make it self-adjoint). It is readily verified that
this operator commutes with $H$. The study of operators of this
type is the basis of the operator approach (sometimes know as the
``evolving constants'' method). The spectrum of the operator
Eq(1.8) is computed from which one may compute a projection
operator $P_{\a}$ say, onto a range of the spectrum. The
associated probability is then of the form ${\rm Tr} ( P_{\a}
\rho)$.

\subhead{\bf 1(C). Decoherent Histories}

In this paper we are primarily concerned with the decoherent
histories approach to quantum theory, and this provides
a second method of calculating probabilities of interest
On the face of it this method is quite different to the operator
method outlined above, and as stated in the Introduction, there
is no particular reason to assume that the methods are equivalent.

In the usual formulation in non-relativistic quantum mechanics
[\cite{GeH,Gri,Omn}],
the central object of interest is the decoherence functional,
$$
D (\a, \a' ) =  {1 \over N} {\rm Tr} \left( \rho_f C_{ \a} \rho
C_{\a'}^{\dag} \right)
\eqno(1.9)
$$
where the histories are characterized by the class
operators $C_{\a}$, which satisfy
$$
\sum_{\a} C_{\a} = 1
\eqno(1.10)
$$
In the simplest case, for non-relativistic systems,
the class operators
are given by time-ordered sequences of
projection operators
$$
C_{\a} = P_{\a_n} (t_n) \cdots
P_{\a_1} (t_1)
\eqno(1.11)
$$
where
$\a$ denotes the string of alternatives $\a_1, \a_2 \cdots
\a_n$. The theory is, however, more general than this
and we will exploit this generality here [\cite{Har3,IsL}]. We have included
the possibility of both an initial state $\rho$ and
a final state $\rho_f$ (normally taken to be proportional
to the identity), and we therefore have to include
the normalization factor $ N =
({\rm Tr} (\rho_f \rho))^{-1} $.

Intuitively, the decoherence functional is a measure of the
interference between pairs of histories $\a$, $\a'$. When  its
real part is zero for all pairs of histories with $\a \ne \a' $,
we say that the histories are consistent and probabilities
$$
p (\a ) = D (\a, \a )
\eqno(1.12)
$$
obeying the usual probability sum rules may be assigned to them.
Typical physical mechanisms which produce this situation usually
cause both the real and imaginary part of $ D (\a, \a') $
to vanish. This condition is usually called decoherence
of histories, and is related to the existence of so-called
generalized records [\cite{GeH,Hal6}].
Note also that when there is decoherence, using
(1.10) the probabilities may be written
$$
p(\a) = {1 \over N } {\rm Tr} \left( \rho_f C_{\a} \rho \right)
\eqno(1.13)
$$

In its application to the parametrized systems considered
here, the initial and final states are attached to the class
operators using the induced inner product scheme [\cite{HaMa}].
So for pure initial and final states, the decoherence functional
is
$$
D(\a, \a') = {1 \over N} \left( \psi_f \circ C_{\a} \circ \psi
\right) \left( \psi_f \circ C_{\a'} \circ \psi \right)^*
\eqno(1.14)
$$
where $\circ$ here denotes the induced inner product.
(We may of course sum over initial or final pure states
to get mixed ones.)
The key question for the models considered here
is then the construction of class operators corresponding
to questions of interest. This is the step analogous to the construction
of operators above in Eq.(1.8), and therefore the class operators must somehow
incorporate reparametrization invariance. It should be stated at this stage
that there does not at present seem to be a completely clear and unambiguous
prescription for constructing the class operators, and part of the
aim of this paper is therefore to explore possible constructions
and examine their properties.

One very natural approach to calculating the class operators
for reparametrization-invariant systems is to use path integrals
[\cite{HaMa,Har3}].
For the relativistic particle
these have the form
$$
C_{\a} (x^{\pp}, x') = \int d T \ g_{\a} (x^{\pp},T|
x',0 )
\eqno(1.15)
$$
where
$$
g_{\a} (x^{\pp},T| x',0 ) = \int_{\a} {\cal D} x^{\mu} \exp \left(
- i \int_0^T ds\left[ { \eta_{\mu \nu} {\dot x}^{\mu} {\dot
x}^{\nu} \over 4} + m^2  \right] \right) \eqno(1.16)
$$
The path integral for $g$ has the form of a non-relativistic propagator.
The sum is over paths from spacetime points $x'$ to $x^{\pp}$
where the paths are restricted in some way defined by
the coarse-graining $\a$. For example, we might be interested
in the probability that the particle passes through some region
of spacetime, or not. More details about the construction of
this object, including the specification of the range
of the $T$ integration, will be given below.

An issue which arises with this definition is that these class
operators often do not everywhere satisfy the constraint equation
with respect to their end-points. As we shall see in particular
examples, they satisfy the constraint except on the boundaries of
the regions defining the coarse grainings. This is an issue
because the induced inner product is only defined between
solutions to the constraint. The auxiliary inner product is still
defined, but in operating with such a class operator on the
initial state, we are stepping out of the constraint surface.
Fortunately, we will find
in particular examples that the problem is easily
fixed by a small amount of intuitively sensible
doctoring on the class operators, guided by the requirement
that path integral methods agree with operator methods.
In particular, we will follow the suggestion of Hartle and
Marolf, which, loosely speaking, is to replace $C_{\a}$
by a new object $C_{\a}'$ which satisfies the constraint
equation everywhere, and satisfies the same essential
boundary conditions as the path integral-defined
object $C_{\a}$ [\cite{HaMa}].

It is easily seen that the operator and decoherent
histories approaches are different, with no guarantee
of their equivalence in general. They both deal with
objects which are compatible with the constraint
(the operator (1.8) and the class operators $C_{\a}$),
but the operator method looks at a projection operator
onto ranges of the spectrum of (1.8), whilst the
decoherent histories approach works with the
class operators $C_{\a}$ and the decoherence functional
(1.14). The class operators are not projections
in general, and in some sense are generalizations of
projections to non-commuting alternatives, so the
two formalisms are quite different. However, it is known
that when there is exact decoherence in the decoherent
histories approach, the probabilities for histories
Eq.(1.12) may be written in the form $ {\rm Tr} ( R_{\a} \rho)
$ where $R_{\a}$ is a projection operator (corresponding
to the existence records mentioned earlier [\cite{GeH,Hal6}]).

\subhead{\bf 1(D). Summary of this Paper}

As stated, the main aim of this paper is to derive expressions
for the probability of crossing a spacelike surface in relativistic
quantum mechanics, using the decoherent histories approach,
and the similar operator approaches.

To get a feel for the formalism for reparametrization-invariant
systems,
we start in Section 2 by applying the formalism to the
non-relativistic particle in parametrized form.
To prepare the way for the study
of the Klein-Gordon equation, we then briefly review some useful
aspects of relativistic quantum mechanics in Section 3.

Our main results are described in Sections 4 and 5 where we apply
the formalism outlined above to the relativistic particle. In
Section 4, we construct a position operator which commutes with
the constraint. Its eigenstates are the Newton-Wigner states, and
in fact the operator is essentially the same as the Newton-Wigner
operator [\cite{NeW}]. The associated probabilities on spacelike surfaces
are those one would anticipate on the basis of the Schr\"odinger
equation which is the square root of the Klein-Gordon equation.
Another, different, candidate expression for the probability
associated with a section of spacelike surface is the flux of the
Klein-Gordon current (with the sign of the negative frequency part
changed, as in Eq.(1.7)). This probability is related to a
different set of position states which are non-orthogonal but
relativistically invariant. There are, therefore, even at this
simple level of canonical quantization, two distinct quantizations
of the relativistic particle, which are not equivalent
(corresponding loosely speaking to ``quantize then constrain''
versus ``constrain then quantize''). We shall refer to them as the
Klein-Gordon (KG) and Newton-Wigner (NW) quantizations.

In Section 5 we consider the decoherent histories analysis of the
system in the KG quantization. We compute the decoherence
functional for histories which cross a surface of constant $x^0$
in a spatial region $\Delta$ (or in its complement $\bar \Delta
$). Since the paths move backwards and forwards in time, the
notion of crossing is ambiguous and needs to be carefully defined.
We show that the essentially unique notions of crossing associated
with the KG quantization are first and last crossing. We thus obtain
probabilities associated with the spacelike surface of the form of
a Klein-Gordon inner product (with a sign change in the negative
frequency sector). However, the histories are only approximately
decoherent (with the off-diagonal terms proportional to the
overlap of the relativistically invariant position states).

The computation of the crossing probabilities hinges on resolving
a subtle point: the path integral representation of the
class operator for not crossing suggests that there is a non-zero
amplitude that the particle will never cross a spacelike surface,
contrary to intuition. This issue turns out in fact to be related
to the problem of class operators which do not satisfy the constraint
mentioned above. An important part of the analysis of Section
5 is a demonstration of how the class operators may be modified
in a sensible way so that they do satisfy the constraint.
The properly modified class operator for not crossing a spacelike
surface then turns out to be zero, in agreement with physical intuition.

In Section 6, we consider the decoherent histories analysis in
the NW quantization. This is much simpler, being very similar in
form to non-relativistic quantum mechanics. Decoherence is exact
and the expect NW probability expressions are easily recovered.

We summarize and conclude in Section 7.

This paper is related to a number of other works in the
field. It exploits and extends the general formalism
of the decoherent histories approach applied to quantum
cosmology set out by Hartle [\cite{Har1}], and more recently
by Hartle and Marolf [\cite{HaMa}]. The application of
this formalism appears to have been applied to particular
models in only two other places. Whelan [\cite{Whe}] considered
probabilities on timelike surfaces for the relativistic
particle (but not using the induced inner product structure,
as here). Craig and Hartle [\cite{CrHa}] have applied the
formalism to a Bianchi IX quantum cosmological model.
There is also some connection with the work on probabilities
for non-trivial spacetime coarse grainings in non-relativistic
quantum mechanics [\cite{YaT}]. A more general
investigation of these ideas applied to quantum cosmological
models is currently being pursued in Ref.[\cite{HaTh}].

\head{\bf 2. The Parametrized Non-Relativistic Particle}

The very first testing ground for ideas about the quantization of
reparametrization invariant
systems is the parametrized non-relativistic particle.
This is the usual non-relativistic particle but with the time
coordinate $t$ raised to the status of a dynamical variable,
with conjugate momentum $p_t$. Its action in Hamiltontian
form is
$$
S = \int ds \left( p_x \dot x + p_t \dot t - N H \right)
\eqno(2.1)
$$
where a denote denotes differentiation with respect to the
parameter $s$. $N$ is a Lagrange multiplier enforcing the
constraint
$$
H = p_t + h = 0
\eqno(2.2)
$$
where $h$ is the usual Hamiltonian $h = p_x^2 / 2m $.
Canonical quantization leads to the Schr\"odinger equation,
$$
H \psi  = \left( p_t + h \right) \psi (x,t) = \left(
- i{ \partial \over \partial t} + h \right) \psi (x,t) = 0
\eqno(2.3)
$$
In terms of dynamics nothing new is gained at this stage.
But the interesting question is to see what the
usual expressions for probabilities look like in the
language introduced in Section 1.

Following the general scheme, we normalize solutions
to the constraint by first considering eigenstates of
$H $, as in Eq.(1.5).
They are normalized using the auxiliary inner product according to
$$
( \Psi_{\l k}, \Psi_{\l' k'} )_A = \int dt dx \ \Psi_{\l k}^* (x,t)
\Psi_{\l' k'} (x,t) =
\delta (\l - \l') \delta (k - k')
\eqno(2.4)
$$
Since $H = p_t + h$,
the solutions to the eigenvalue equation may be written
$$
\Psi_{\l k} (x,t) = {  1 \over (2 \pi)^{\half} } \ e^{i \l t }
\psi_k (x,t)
\eqno(2.5)
$$
where $\psi_k (x,t)$ satisfies the Schr\"odinger equation.
It follows that
$$
{ 1 \over 2 \pi}
\int dt \int dx \ e^{ - i \l t + i \l't }
\ \psi^*_k (x,t) \psi_{k'} (x,t) = \delta ( \l - \l' ) \delta (k-k')
\eqno(2.6)
$$
The integral contains within it the usual inner product
$$
(\psi_k, \psi_{k'} )_S =
\int dx \ \psi^*_k (x,t) \psi_{k'} (x,t)
\eqno(2.7)
$$
This has the important property that it
is independent of time when the states obey
the Schr\"odinger equation, so the time integral may be done in
Eq.(2.6), pulling down a delta function $\delta (\l - \l')$,
and it follows that
$$
(\psi_k, \psi_{k'} )_S = \delta (k-k')
\eqno(2.8)
$$
This means that the expected Schr\"odinger inner product
on surfaces of constant $t$ follows from the induced inner product
defined on the whole of spacetime.

We may now ask for the probabilities in various situations
of interest. Perhaps the simplest is the probability
of finding the particle in the spatial region $\Delta$ at time $t_0$.
In the usual approach it is of course
$$
p_{\Delta} = \int_\Delta dx \ | \psi (x,t_0) |^2
\eqno(2.9)
$$
To express this in the language of Section 1,
we seek an operator which commutes with the constraint
$H$ and corresponds to the answer to ``the value
of $x$ when $t = t_0$''. Following the general scheme
this is
$$
X = \int_{-\infty}^{\infty}
ds \ {d t (s) \over ds} \ x (s) \ \delta ( t(s) - t_0)
\eqno(2.10)
$$
where $x(s)$ and $t(s)$ are the evolution of $x$ and
$t$ using the constraint $H$ as a Hamiltonian:
$$
x(s) = e^{ i H s } x e^{ - i H s },
\quad
t(s) = e^{ i H s } t e^{ - i H s }
\eqno(2.11)
$$
Since $H = p_t + h $ this is
$$
\eqalignno{
x(s) &= e^{ i h s } x e^{ - i h s } = x + { p s \over m }
&(2.12) \cr
t(s) &= e^{ i p_t s} t e^{ -i p_t s } = t + s
&(2.13) \cr }
$$
The integral over $s$ may be done in Eq.(2.10) with the result
$$
X = x - { p (t -t_0) \over m}
\eqno(2.14)
$$
It is easy to confirm that this commutes with $H$.

Since $H$ and $X$ commute, they possess a joint set of eigenstates
$ u_{\l \bar x} $.
The eigenvalue equation for $X$ is
$$
X u_{\l \bar x}(x,t) = \bar x u_{\l \bar x} (x,t)
\eqno(2.15)
$$
with solutions
$$
u_{\l \bar x} (x,t) = { 1 \over (2 \pi)^{\half} }
e^{ i \l t} g (x,t | \bar x, t_0 )
\eqno(2.16)
$$
where $g$ is the non-relativistic propagator. In the auxiliary
inner product these are normalized according to
$$
( u_{\l \bar x}, u_{\l' \bar x'} )_A = \delta (\l - \l')
\ \delta (\bar x - \bar x')
\eqno(2.17)
$$

The amplitude for an eigenstate of the constraint of
the form
$$
\Psi_{\l'} (x,t) = { 1 \over (2 \pi)^\half} e^{ i \l t} \psi (x,t)
\eqno(2.18)
$$
to be in an eigenstate of $X$ is
$$
\eqalignno{
( u_{\l \bar x}, \Psi_{\l'} )_A
&= {1 \over (2 \pi) }
\int dt dx \ e^{ i t ( \l - \l') }
g^* (x,t |\bar x, t_0) \psi (x,t)
\cr
&= \delta (\l - \l') \psi (\bar x, t_0)
&(2.19) \cr}
$$
The probability is then computed from the expression
$$
\int_{\Delta} d \bar x
\ ( \Psi_{\l^{\pp}}, u_{\l \bar x} )_A
\ ( u_{\l \bar x}, \Psi_{\l'} )_A
= \delta (\l^{\pp} - \l) \delta( \l - \l')
\int_{\Delta} d \bar x \left| \psi (\bar x, t_0) \right|^2
\eqno(2.20)
$$
Following the induced inner product prescription, we drop
the delta-functions on the right, thereby obtaining
the expect result for the probability, Eq.(2.9).

The operator formalism  with Eq.(1.8) allows one to ask a
richer variety of questions than those normally considered
in non-relativistic quantum mechanics. We may
consider, for example, the question, ``What is the value of $t$ at a given
value of $x$''? The associated operator
is
$$
T = t_0 + { m (x-x_0) \over p}
\eqno(2.21)
$$
A suitable operator ordering must be chosen, but
the presence of the $1/p$ factor makes it difficult
to turn this into a self-adjoint operator (see Ref.[\cite{GRT}],
for example). This operator arises in relation to the arrival time problem in
non-relativistic quantum mechanics, an issue that has
attracted a lot of recent attention in the literature [\cite{Time}].

Both of the above questions in non-relativistic quantum mechanics
may also be analysed using the decoherent histories approach.
We will not go into the details here, except to make some simple
observations that are related to the relativistic particle
case we consider later.

In the decoherent histories approach, the probability Eq.(2.9) may
also be obtained using a standard non-relativistic path integral,
in which one sums over paths which cross the surface $t=t_0$ in
the spatial range $\Delta$. It is a property of this path integral
that the paths cross this surface once and only once, and as a
consequence of this, the histories are exactly decoherent.
As described in Section 1(C), exact
decoherence of histories implies that records exist [\cite{GeH,Hal6}], or
in other words, that the probability may be
written in the form ${\rm Tr} (R_{\a} \rho) $
for some projection operator $R_{\a}$. This is thoroughly consistent
with the existence of the self-adjoint operator (2.14) from which
the probabilities Eq.(2.9) are derived in the operator approach.

But now consider, by contrast, the probability for crossing a
surface of constant $x$. In the decoherent histories analysis of
this question (which is rather non-trivial [\cite{YaT}]), the
paths may cross the surface many times. Furthermore, it is found
that the histories are typically not decoherent (unless an
environment to produce decoherence is included, but we do not
consider that case here), and this appears to be related to the
multiple crossings. We cannot therefore deduce the existence of
records and a probability of the form ${\rm Tr} ( R_{\a} \rho ) $.
There would be an inconsistency with the operator approach here if
there was a self-adjoint operator corresponding to this question.
But interestingly, as we have seen, the corresponding operator
Eq.(2.21) is not self-adjoint. The point therefore, is that
multiple surface crossings and the associated lack of decoherence
in the decoherent histories approach appear to be related to the
absence of a self-adjoint operator in the operator approach. We
will see more evidence of this in the case of the relativistic
particle in Sections 5 and 7.

\head{\bf 3. Green Functions of the Klein-Gordon equation}

The Klein-Gordon equation has a variety of associated
Green functions and it will be useful to briefly
summarize them here. In order to agree with the notation
of Ref.[\cite{HaOr}] (which we follow very closely), we use
particle physics convention in which the signature of
the metric is $(+---)$.
The positive and negative frequency Wightman functions
$G^{\pm}$ are defined by
$$
G^{\pm}(x, y)={1\over (2\pi)^3}\int_{k_0=\pm\om_\k}
{d^3\k \over 2\om_\k}e^{-ik \cdot (x-y)}
\eqno(3.1)
$$
where $\om_k = \sqrt{\k^2 +m^2}$. They satisfy the
composition laws
$$
G^{\pm} = \pm  G^{\pm} \circ G^{\pm},
\quad G^{\pm} \circ G^{\mp}  = 0
\eqno(3.2)
$$
where here and in the remainder of this section $\circ$ denotes the
Klein-Gordon inner product (unless explicitly denoted otherwise).
The causal Green function is defined by
$$
i G(x,y) = G^+ (x,y) - G^- (x,y)
\eqno(3.3)
$$
Its main property is that it propagates all solutions
to the Klein-Gordon equation
$$
\phi = i G \circ \phi
\eqno(3.4)
$$
It also obeys the composition law
$$
G = i G \circ G
\eqno(3.5)
$$
The Hadamard
function is defined by
$$
G^{(1)} (x,y) = G^+(x,y) + G^-(x,y)
\eqno(3.6)
$$
and obeys the composition laws,
$$
G^{(1)} = i G \circ G^{(1)} = i G^{(1)} \circ G,
\quad \quad G  = -i G^{(1)} \circ G^{(1)}
\eqno(3.7)
$$
All of the above are solutions to the Klein-Gordon equation.

The Feynman Green function is
$$
i G_F (x,y) = \theta (x^0 - y^0) G^+ (x,y)
+ \theta (y^0 -x^0) G^- (x,y)
\eqno(3.8)
$$
and satisfies
$$
\left( \square + m^2 \right) G_F(x,y) = - \delta^{(4)} (x-y)
\eqno(3.9)
$$
It obeys the composition laws
$$
G_F = i G_F \circ G_F
\eqno(3.10)
$$

Also of interest is the Newton-Wigner propagator
$$
G_{NW} ( \x, x^0, \y, y^0 )
= {1 \over (2 \pi)^3 } \int_{k_0 = \om_\k}
d^3 \k \ e^{ - i k \cdot (x-y) }
\eqno(3.11)
$$
which is the propagator associated with the positive
square root of the Klein-Gordon equation
$$
i { \partial \phi \over \partial x^0 }
= h \phi
\eqno(3.12)
$$
where $h = \sqrt{ - \nabla^2 + m^2} $.
It is also useful to define a negative frequency Newton-Wigner
propagator, given by (3.11) but with $ k_0 = - \om_k$,
and this will be denoted $\tilde G_{NW}$.
It is easily seen that the Newton-Wigner propagator is
related to the Wightman function by
$$
G_{NW} (x,y) = 2i{ \partial \over \partial x^0} G^+ (x,y) = - 2i{
\partial \over \partial y^0} G^+ (x,y)
\eqno(3.13)
$$

Some of these Green functions can be obtained from a path integral
of the form (1.15), (1.16). An unrestricted sum with $T$
integrated over an infinite range yields the Hadamard function
$G^{(1)}$. (See Fig(3.1)). A half-infinite range, $ 0 \le T <
\infty $, yields $i G_F$, where $G_F$ is the Feynman Green
function. (See, for example, Ref.[\cite{Hal3}]). The Newton-Wigner
propagator can also be obtained from (1.15), (1.16) by summing
over all paths from $y$ to $x$ which never cross the surface of
constant $x^0$, except when they end at the point $x$. (See
Fig.(3.2)). More details of this construction are discussed in
Sections 5 and 6. (See also Ref.[\cite{HaOr}]).

From the path integral representations, one can see that
$G^{(1)}$ corresponds to the operator $\delta (H)$, which is
essentially the identity on the constraint subspace (and so
we effectively have
$\delta (H) | \phi \ra = | \phi \ra $ for solutions to the
constraint). This is perhaps confusing since $G^{(1)}$ does not in
fact propagate positive and negative frequency solutions to the
Klein-Gordon equation (it is the causal Green function $G$ that
does this job, via Eq.(3.4)). The resolution of this is the choice
of inner product. $G^{(1)}$ {\it does} in fact propagate all
solutions if they are attached with the induced inner product
(1.7). For suppose we have a solution $\phi= \phi^+ + \phi^-$.
Then
$$
\eqalignno{
G^{(1)} \circ_I \phi &= (G^+ + G^-) \circ_I ( \phi^+ + \phi^-)
\cr
& =G^+ \circ_{KG} \phi^+ - G^- \circ_{KG} \phi^-
\cr
& = (G^+ - G^-) \circ_{KG} ( \phi^+ + \phi^-)
\cr
&=  i G \circ_{KG} \phi
&(3.14) \cr}
$$
In this sense, $G^{(1)}$ is effectively equivalent to $G$.

It is also interesting note in this connection that it was claimed
in Ref.[\cite{HaOr}] that there is no path integral of the form (1.15), (1.16)
that will yield the causal propagator $G$ directly. Whilst this
is still in some sense true, one can see that it depends on how
the initial states are attached: the path integral for $G^{(1)}$
but with initial states attached using the induced inner product
does in fact effectively give the causal propagator $G$.

We may now consider the form of the decoherence functional
for the Klein-Gordon equation (we follow the construction
of Ref.[\cite{HaMa}]).
We take it to be of the
form (1.14). We take a fixed pure initial state and sum
over a complete set of final states.
This gives
$$
\eqalignno{
D (\a, \a') &= \sum_{\psi_f} ( \psi_f \circ_I C_{\a} \circ_I \psi)
( \psi_f \circ_I C_{\a'} \circ_I \psi)^*
&(3.15) \cr}
$$
where note here that we use the induced inner product. (The
normalization factor is unity in this case).
Since $\psi_f$ denotes a complete set of positive and negative
frequency solutions, it is easy to show that
$$
\sum_{\psi_f}  \psi_f^* (x) \psi_f (y)
= G^{(1)} (x,y)
\eqno(3.16)
$$
Furthermore,  since $C_{\a}$ are solutions to the constraints,
the action of $G^{(1)}$ changes nothing, $G^{(1)} \circ_I C_{\a}
= C_{\a}$, so we have
$$
D(\a, \a') = \psi^* \circ_I C^{\dag}_{\a'} \circ_I C_{\a} \circ_I \psi
\eqno(3.17)
$$
Finally, when there is exact decoherence, $D(\a, \a') = 0$
for $\a \ne \a'$, the probabilities are
$$
\eqalignno{
p(\a) &= D(\a,\a') = \sum_{\a'} D(\a,\a')
\cr
&= \psi^* \circ_I C_{\a} \circ_I \psi
&(3.18)
\cr}
$$
In the KG quantization, the induced inner product becomes
the modified KG inner product (1.7), with initial states normalized
in this inner product. In the NW quantization, states obey
the Schr\"odinger equation, they are normalized in a Schr\"odinger
inner product and $\circ$ is taken to be that inner product
in the decoherence fucntional.

\head{\bf 4. An Operator Approach for the Klein-Gordon Equation}

We now describe the use of operator methods to
obtained probabilities associated with the Klein-Gordon equation.
The relativistic particle is described by the constraint
$$
H = p_0^2 - \p^2 - m^2 = 0
\eqno(4.1)
$$
where the canonical variables $x^{\mu}$, $p^{\nu}$ obey
the commutation relations
$$
[x^{\mu}, p^{\nu} ] = - i \eta^{\mu \nu}
\eqno(4.2)
$$
We are interested in the question, ``What is the
value of $x^k$ when $x^0 =\tau$?''. As indicated
already, there are potentially many ways of formulating
and answering this question. We will first use the operator
methods of Section 1(B). Following Eq.(1.8),
the operator expressing
this question is
$$
X^k = \int_{-\infty}^{\infty}
ds \ \half \{ {d x^0 (s) \over ds} , \delta ( x^0(s) - \tau ) \} \ x^k (s)
\eqno(4.3)
$$
where $\{ \ , \ \} $ denotes the anticommutator, and we have
$$
\eqalignno{
x^0 (s) &= x^0 + p^0 s
&(4.4) \cr
x^i (s) &= x^i + p^i s
&(4.5) \cr }
$$
The object of interest is therefore given by
$$
X^k = x^k - {p^k \over 2}   \{ {1 \over p^0}, (x^0 -\tau ) \}
\eqno(4.6)
$$
and it is easily seen that this commutes with $H$.
One might anticipate that the $1/p^0$ factor may present
problems in turning this into a self-adjoint operator, but
this problem does not arise since we are looking
for eigenstates of $X^i$ which also satisfy the constraint,
and this bounds $p_0$ away from zero. (Essentially the same operator
was also considered by Marolf [\cite{Mar1}]).

We choose a momentum representation, in which
$$
x^k \ria - i { \partial \over \partial p_k},
\quad x^0 \ria -i{ \partial \over \partial p^0}
\eqno(4.7)
$$
and $X^k$ is
$$
X^k = \left( - i { \partial \over \partial p_k}
+ i{ p^k \over p^0} { \partial \over \partial p^0}
- i { p^k \over 2 (p^0)^2}
+ {p^k \tau \over p^0} \right)
\eqno(4.8)
$$
(recalling that $p^k = - p_k$ with our choice of signature).
This is self-adjoint in the (momentum space version of)
the auxiliary inner product Eq.(1.4).
The eigenstates of $X^k$ are the functions
$$
f(p) = { 1 \over (2 \pi)^{3/2} }
( 2 p^0)^{\half} \ e^{ i p^0 \tau - i  \p \cdot \x}
\ g( p \cdot p  )
\eqno(4.9)
$$
where $g$ is any function of $p \cdot p$, and the
eigenvalue is $x^k$. For these to
be eigenstates of the constraint we also need to choose
$ g  = \delta ( p \cdot p - m^2  ) $. Introducing the
eigenstates $ | p \ra $ of $ p^{\mu}$, (where
$ \la p | p' \ra = \delta^{(4)} (p-p')$), the eigenstates
of $X^i$ may be written
$$
\eqalignno{
| \x, \tau \ra &= { 1 \over (2 \pi)^{3/2} } \int d^4 p
\ (2 p^0)^{\half} \ e^{ i p^0 \tau - i  \p \cdot \x}
\ \delta (p^2 - m^2  ) \ | p \ra
\cr
&= { 1 \over (2 \pi)^{3/2} }
\int { d^3 p \over (2 \om_p)^{\half} }
\ e^{ i \om_p \tau - i \p \cdot \x } \ | \p +\ra
+ { i \over (2 \pi)^{3/2} }
\int { d^3 p \over (2 \om_p)^{\half} }
\ e^{ - i \om_p \tau - i \p \cdot \x } \ | \p -\ra
\cr
& = | \x, \tau + \ra \ +  \ i | \x, \tau - \ra
&(4.10) \cr}
$$
(The factor of $i$ in the negative frequency term, not present
in other definitions of these states [\cite{HaKu,HaOr}], does not in
fact make any difference.)
Here, we have introduced the momentum states $| \p \pm \ra $
on the positive and negative frequency sectors
which are normalized according to
$$
\la \p \pm | \p' \pm\ra = 2 \om_\p \delta (\p - \p')
\eqno(4.11)
$$
and also $ \la \p \pm | \p' \mp \ra =0$.
The states $ | \x, \tau \ra $ are the
Newton-Wigner states [\cite{NeW}]. They are orthogonal
at equal times, and satisfy the completeness relation
$$
1 = \int d^3x \ | \x, \tau +\ra \la \x,
\tau +| + \int d^3x \ | \x, \tau  - \ra \la \x, \tau -|
\eqno(4.12)
$$
The probability of entering a region $\Delta$ at time $\tau$
is given by
$$
p_{\Delta} = \int_{\Delta} d^3 x \ \left| \la \x, \tau +| \psi \ra
\right |^2
+ \int_{\Delta} d^3 x \ \left| \la \x, \tau -| \psi \ra
\right |^2
\eqno(4.13)
$$
where the states $ \psi_{\pm} (\x, \tau ) = \la \x, \tau \pm | \psi \ra$
obey the Klein-Gordon equation and its positive/negative square
root, and are normalized in a Schr\"odinger inner product.

The Newton-Wigner states could also have been obtained by solving
the constraint classically and then considering the eigenstates
of the operators
$$
X^k = x^k \pm { p^k \tau \over \sqrt{ \p^2 +m^2} }
\eqno(4.14)
$$
The Newton-Wigner states therefore
correspond to ``constraining before quantization''.
It is also important to compare with the position operator
introduced by Newton and Wigner [\cite{NeW}], which is, on the surface
$x^0 = 0$, in the momentum
representation
$$
X^k_{NW} = -i { \partial \over \partial p_k} - {p^k \over
2 \om_{\p}^2 }
\eqno(4.15)
$$
This is in fact the same as the operator
$$
\int d^3 x \ | \x, \tau + \ra \ x^k \ \la \x, \tau + |
\eqno(4.16)
$$
with $\tau = 0$, in terms of the Newton-Wigner states above.
Eq.(4.15) is not the same as (4.6), since the constraint holds
in Eq.(4.15), but has not yet been imposed in Eq.(4.6).
Eq.(4.15) is, however, the same as Eq.(4.14), with $\tau =0$,
once one recognizes that the inner product structure (4.11)
requires the replacement
$$
x^k \ \ria \ - (2 \om_{\p})^{\half} \ i { \partial \over \partial
p_k} \ { 1 \over ( 2 \om_{\p})^{\half} } = -i { \partial \over
\partial p_k} - {p^k \over 2 \om_{\p}^2 } \eqno(4.17)
$$
There is therefore agreement with the earlier work of
Newton and Wigner.

An alternative way of defining position states is to first
consider eigenstates of the position operator $ \hat x^{\mu}$ on the
auxiliary Hilbert space, and then project onto the
constraint subspace using $\delta (H)$. This corresponds
to quantizing before constraining, and yields
$$
\eqalignno{
| x  \ra
&= { 1 \over (2 \pi)^{3/2} }
\int_{p_0 = \om_{\p}} { d^3 p \over 2 \om_p }
\ e^{ i p \cdot x } \ | \p +\ra
+ { 1 \over (2 \pi)^{3/2} }
\int_{p_0 = - \om_{\p}} { d^3 p \over 2 \om_p }
\ e^{ i p \cdot x } \ | \p -\ra
\cr
&= | x+ \ra \ + \ | x - \ra
&(4.18) \cr}
$$
Unlike the Newton-Wigner states, these states are Lorenz-invariant.
Furthermore,
they not orthogonal, since
$$
\la x | y \ra = G^{(1)} (x,y)
\eqno(4.19)
$$
although they are approximately orthogonal in the sense that
$ G^{(1)} (x,y)$ decays when $x$ and $y$ are separated by more
than the Compton wavelength $m^{-1}$.
They also obey a completeness relation
$$
1 = i \int d^3 x \ \left( | x+\ra \ulra{\partial_0} \la x+|
\ -
\ | x-\ra \ulra{\partial_0} \la x-| \right)
\eqno(4.20)
$$
These properties of the states $ | x \ra $ are reminiscent
of the coherent states, and suggest
that the probability for crossing a spacelike surface
$x^0 = \tau $ in the region $\Delta $ may be taken to be
$$
p_{\Delta} = i \int_{\Delta} d^3 x \ \left( \phi^*_+ \ulra{\partial_0}
\phi_+ \ - \ \phi^*_- \ulra{\partial_0} \phi_-
\right)
\eqno(4.21)
$$
The states $\phi_{\pm} (x) = \la x \pm | \phi \ra $ are positive/negative
frequency solutions to the Klein-Gordon equation.
The minus sign in the second term ensures that the expression is
positive, and in the limit $ \Delta = \R^3 $ this expression
becomes the norm of $\phi$ in the induced inner product, as required.

\head{\bf 5. Decoherent Histories Analysis in the Klein-Gordon
Quantization}

We now come to the main point of this paper which is
to use the decoherent histories approach to compute
the answer to the question, ``What is the probability
that the particle is found in the spatial region $\Delta$ at time
$x^0=\tau$?''. In this Section, we consider the KG quantization,
with the aim of obtaining Eq.(4.21), and we consider the NW
quantization and Eq.(4.13) in Section 6.

The decoherence functional is given by Eq.(3.17).
We take a pure initial state, and we sum over
a complete set of positive and negative solutions in
the final state. The main issue is to compute the
class operators $C_{\a} (x^{\pp},x')$ corresponding to crossing $x^0 = \tau$
either inside the region $\Delta$ or outside it, in
its complement $\bar \Delta$. We expect that these
operators can be obtained by a sum over paths
which either cross or do not cross the region (in a sense
to be made more precise below). However, in order
to tackle a subtlety
first noted in Section 1 concerning the definition of
the class operators, we first need to consider a simpler question.

\subhead{\bf 5(A). The Class Operator for Not Crossing a Spacelike
Surface}

In computing a class operator of the form $C_{\a}(x^{\pp},x')$
we sum over paths from $x'$ to $x^{\pp}$ satisfying
some condition specified by the coarse graining $\a$.
However, this construction appears to allow for the possibility
of defining a coarse graining consisting of paths from
$x'$ to $x^{\pp}$ which never cross a given spacelike surface.
(See Fig.(5.1)).
This is at first sight disconcerting. Every classical trajectory
of the relativistic particle crosses every spacelike surface
(unless it is tachyonic). And in the quantum theory, a solution
to the Klein-Gordon equation cannot be zero on one side of
a spacelike surface and non-zero on the other.

To investigate this issue,
we consider the following coarse graining:
paths from $x'$ to $x^{\pp}$ which either cross or do not cross
the surface $x^0 = \tau$. That is, we are interested in the
question, given a solution to the Klein-Gordon equation,
what is the probability that the particle will be found
on the spacelike surface $x^0 = \tau$, or will never be
found on that surface? Clearly the answer
to this question must be probability unity for crossing
the surface, and probability zero for not crossing.

The sum over paths which do not cross $x^0=\tau$ is easily constructed
using the path integral representations(1.15), (1.16). We take
$T$ to have an infinite range and we sum over all paths which never
cross the surface. In Ref.[\cite{HaOr}] it was
shown that this is in fact equivalent to a method of images
construction for $g(x^{\pp}, T |x', 0 )$,
and we obtain, for the class operator for paths restricted
to not cross the surface,
$$
\eqalignno{
C_r (x^{\pp},x') = & \left( \theta ({x^0}^{\pp}-\tau) \theta ({x^0}'-\tau)
+ \theta (\tau-{x^0}^{\pp}) \theta (\tau-{x^0}') \right)
\cr
& \times \left( G^{(1)} (x^{\pp},x') - G^{(1)} (\tilde x^{\pp}, x') \right)
&(5.1)\cr}
$$
where $\tilde x $ denotes the reflection of the point $x$ about
$x^0 = \tau$, that is,
$$
\tilde x^{\mu} = (2\tau -x^0, \x )
\eqno(5.2)
$$
$C_r$ vanishes on ${x^0}' = \tau$ and ${x^0}^{\pp} = \tau$.
Both $G^{(1)} (x^{\pp},x')$ and
$G^{(1)} (\tilde x^{\pp},x')$ are solutions to the KG equation, but
the presence of the $\theta$-functions means that $C_r$
does not satisfy it,
$$
( \square +m^2) C_r (x^{\pp},x') = 2 \delta ({x^0}^{\pp} - \tau)
\ \e ({x^0}' -\tau ) \ \partial_0 G^{(1)} (\tau, \x^{\pp} | {x^0}', \x')
\eqno(5.3)
$$
where $\e(x)$ is the signum function. This is the difficulty
with the path integral-defined class operators mentioned
in Section 1.
The sum over paths which always cross $x^0=\tau$, which we denote $C_c(x^{\pp},x')$,
is constructed using the relation
$$
G^{(1)}(x^{\pp},x')  = C_c(x^{\pp},x') + C_r (x^{\pp},x')
\eqno(5.4)
$$
from which one can see that $C_c(x^{\pp},x') $ will also not satisfy the constraint
(although to actually compute the crossing class operator we will use
a different method below).

Since the constraint
is associated with reparametrization invariance, it might
seem that failure to satisfy it is associated with a breaking
of that invariance. However, the connection between constraints
and invariance can be rather subtle for the case of reparametrizations.
In particular, the path integral (1.15), (1.16) between
fixed end-points over paths restricted to pass through certain
spacetime regions is clearly reparametrization invariant.
These issues are discussed more fully in Refs.[\cite{Har3,HalH}].

Note also that we have now encountered {\it two} potential
difficulties which are not obviously related: firstly,
the possibility of a physically unreasonable
result, and secondly, the fact that the naturally defined
class operator does not satisfy the constraint. We will
now see that both of these problems are solved simultaneously
by appropriate modification of the class operator.

Following the suggestion of Hartle and Marolf [\cite{HaMa}],
we will deal with this issue
by replacing $C_r$ by another object $ C_r'$ which satisfies
the most important boundary conditions defining $C_r$ but
which also satisfies the constraint everywhere. The
boundary conditions satisfied by $C_r$ are that, (a), it vanishes
on ${x^0}^{\pp}=\tau$, ${x^0}'=\tau$ and when $x^{\pp}$ and $x'$ are on opposite
sides of the surface. One might also be tempted to impose that,
(b), $C_r'$ coincides with (the non-zero) $C_r$ when $x^{\pp}$ and $x'$ are
on the same side of the surface. It appears to be impossible
for a function satisfying the constraint everywhere
to satisfy all of these conditions. The essence of the non-crossing
propagator appears to be contained in the conditions (a),
so we drop conditions (b). The unique solution to the
constraint equation satisfying conditions (a) is then
quite simply
$$
C_r' (x^{\pp},x') = 0
\eqno(5.5)
$$
This is because for fixed $y$, $C_r'(x^{\pp},x')$ is zero for all
values of $x^{\pp}$ on the opposite side of the surface to $x'$.
This means that both $C_r'$ and its normal derivative are
zero on all spacelike surfaces on the opposite side of the
surface $x^0= \tau$, and the solution to the Klein-Gordon
equation with these conditions is simply $C_r' = 0$.
The only way of getting a non-zero results as $x^{\pp}$ moves
from the opposite side to the same side as $x'$ is to have
a discontinuity in the normal derivative, but this can only
be achieved with a delta-function source, as in Eq.(5.1).

The conclusion (5.5) for the modified
class operator for non-crossing paths implies that the
modified class operator for paths that always cross
is quite simply $G^{(1)} (x,y) $. It now follows
that histories partitioned according to whether or not
they cross $x^0 = \tau $ are exactly decoherent,
the probability of not crossing is zero, and the
probability of crossing is $1$.
We have therefore shown that by sufficiently careful
treatment of the class operators and their boundary conditions,
we obtain the expected and physically sensible result.

\subhead{\bf 5(B). Crossing Propagators and the Path Decomposition
Expansion}

Having resolved the issue of how to modify class operators that do
not satisfy the constraint, we may now turn to the issue of
computing the class operators for crossing the spacelike surface
$x^0 = \tau$ in
a spatial region $\Delta$. These operators (before modification)
will be constructed by a path integral of the form (1.15), (1.16)
in which the paths cross the spacelike surface in the region
$\Delta$. Because the paths go backwards and forwards in time,
they will typically cross a given spacelike surface many times, so
the notion of crossing needs to be specified more precisely.
We will see however, that for the path integral representations
of the Klein-Gordon propagators, first and last crossings
are in fact the only useful notions of crossing.

A very useful result for our purposes is the path decomposition
expansion, or PDX [\cite{PDX,HaOr}]. For a propagator of the non-relativistic
form (1.16), it implies that when $x^{\pp}$ and $x'$ are on
opposite sides of the surface of constant $x^0$, we have
$$
g (x^{\pp}, T | x', 0 ) = 2i \ \e (\tau - {x^0}')  \ \int_0^T dt_c \int d^3 x
\ g(x^{\pp},T | x, t_c) \  \ura{\partial_0} g (x, t_c | x', 0 )
\eqno(5.6)
$$
This formula is obtained by partitioning the paths in the
sum over histories (1.16) according to the parameter time $t_c$
and position ${\bf x}$ at which they cross the surface of constant
$x^0$ for the first time\footnote{$^{\dag}$}{The full version of the PDX
actually involves the normal derivative of the restricted
propagator $g_r$, but in this simple case, $g_r$ may be
computed using the method of images and it
follows that ${\partial_0} g_r = 2 {\partial_0} g$,
which is what is used in Eq.(5.6). See Ref.[\cite{HaOr}] for
more details.}. (See Fig.(5.2)).
If we partition according to the
{\it last} crossing, we get
$$
g (x^{\pp}, T | x', 0 ) = - 2i\  \e ( {x^0}^{\pp} - \tau )
\ \int_0^T dt_c
\int d^3 x  \ g(x^{\pp},T | x, t_c)
 \ula{\partial_0} g (x, t_c | x', 0 )
\eqno(5.7)
$$
When $x^{\pp}$ and $x'$ are on the same side of the surface,
there is the possibility of paths between these points which
do not cross the surface.
The appropriate formulae then are
$$
\eqalignno{
g (x^{\pp}, T | x', 0 ) =  & \ g_r (x^{\pp}, T| x', 0 )
\cr
& + 2i \ \e (\tau - {x^0}') \ \int_0^T dt_c
\int d^3 x  \ g(x^{\pp},T | x, t_c)
\  \ura{\partial_0} g (x, t_c | x', 0 )
&(5.8)\cr }
$$
for the first crossing and
$$
\eqalignno{
g (x^{\pp}, T | x', 0 ) = &\ \ g_r (x^{\pp}, T| x', 0 )
\cr
& - 2i \ \e ( {x^0}^{\pp} - \tau )  \ \int_0^T dt_c
\int d^3 x  \ g(x^{\pp},T | x, t_c)
\ \ula{\partial_0} g (x, t_c | x', 0 )
&(5.9) \cr}
$$
for the last crossing, where $g_r$ denotes the restricted propagator
given by a sum over paths which never cross the surface.
The formulae (5.6)--(5.9) imply that the sum over paths which cross
the surface is given by either (5.6) or (5.7),
irrespective of whether $x'$ and $x^{\pp}$ are on the same side
or opposite sides of the surface. (But only in the latter case
are these expressions then equal to the full propagator).

These results were used in Ref.[\cite{HaOr}] to derive the composition laws
of relativistic propagators from the path integral. Here, we note
that the propagators for first or last crossing the surface
$x^0 = \tau$ in the region
$\Delta$ are readily obtained by simply restricting the $d^3 x$
integration to the region $\Delta$.

\subhead{\bf 5(C). First and Last Crossing Relativistic Propagators}

Turning now to the relativistic propagators, the class operator
for crossing the surface $x^0 = \tau $ in the region
$\Delta $ is
$$
C_{\Delta}(x^{\pp},x') = \int_{-\infty}^{\infty} d T \ g_{\Delta}
(x^{\pp}, T | x', 0 )
\eqno(5.10)
$$
We first take the case
where $g_{\Delta} (x^{\pp}, T,x',0)$ is the sum over paths
from $x'$ to $x^{\pp}$ in fixed
proper time $T$ which cross the surface for the first time
in $\Delta$, that is,
$$
g_{\Delta} (x^{\pp}, T | x', 0 ) = 2i \ \e (\tau - {x^0}')\ \int_0^T dt_c
\int_{\Delta} d^3 x  \ g(x^{\pp},T | x, t_c)
\  \ura{\partial_0} g (x, t_c | x', 0 )
\eqno(5.11)
$$
(See Fig.(5.3)).
This is valid for $x^{\pp},x'$ on either the same or opposite
sides of the spacelike surface.
Inserting in Eq.(5.10), writing the integral as a sum of
two parts corresponding to the positive and negative
ranges of $T$,
and changing variables to $v = T - t_c $,
$u = t$ (see Ref.[\cite{HaOr}] for more details),
this yields,
$$
C^f_{\Delta} (x^{\pp},x') = -2 i\ \e (\tau - {x^0}')\ \int_{\Delta} d^3 x
\ \left[ G_F (x^{\pp}, x) \ura{\partial_0} G_F (x,x')
- G_F^* (x^{\pp}, x) \ura{\partial_0} G_F^* (x,x') \right]
\eqno(5.12)
$$
This is the formula for first crossing the region
$\Delta$ for all end-points. It is also conveniently written,
$$
C^f_{\Delta} (x^{\pp}, x') = -
\int_{\Delta} d^3 x \ \left[ G^{(1)} (x^{\pp},x) \ura{\partial_0}
G (x,x') + \e ( {x^0}^{\pp} - \tau ) \e ( \tau - {x^0}')
\ G(x^{\pp},x) \ura{\partial_0} G^{(1)} (x,x') \right]
\eqno(5.13)
$$
It is also of interest to consider the class operator
defined by the last crossing, which is easily shown to be
$$
\eqalignno{
C^{\ell}_{\Delta} (x^{\pp},x') =& 2 i\ \e ({x^0}^{\pp}-\tau )
\ \int_{\Delta} d^3 x
\ \left[ G_F (x^{\pp}, x) \ula{\partial_0} G_F (x,x')
- G_F^* (x^{\pp}, x) \ula{\partial_0} G_F^* (x,x') \right]
\cr
=& \int_{\Delta} d^3 x \ \left[ \e ( {x^0}^{\pp} - \tau )
\e ( \tau - {x^0}')
G^{(1)} (x^{\pp},x) \ula{\partial_0}
G (x,x') \right.
\cr
& \quad \quad \quad \quad \quad  +
\left. \ G(x^{\pp},x) \ula{\partial_0} G^{(1)} (x,x') \right]
&(5.14)\cr}
$$
When the initial and final points are on opposite sides of the surface,
we have
$$
\e ( {x^0}^{\pp} - \tau ) \e ( \tau - {x^0}') =1
\eqno(5.15)
$$
It is then convenient to average the first and last crossing class operators
to obtain
$$
\eqalignno{
C_{\Delta}( x^{\pp}, x') &= \half \left( C^f_{\Delta}( x^{\pp}, x')
+ C^{\ell}_{\Delta}( x^{\pp}, x') \right)
\cr
&= - \half
\int_{\Delta} d^3 x \left[G^{(1)} (x^{\pp},x) \ulra{\partial_0}
G (x,x') +
\ G(x^{\pp},x) \ulra{\partial_0} G^{(1)} (x,x') \right]
\cr
&= i \int_{\Delta} d^3 x
\ \left( G^+ (x^{\pp},x) \ulra{\partial_0} G^+ (x,x')
-G^- (x^{\pp},x) \ulra{\partial_0} G^- (x,x') \right)
&(5.16) \cr}
$$
It is then readily confirmed, using the properties
Eq.(3.2) and (3.7), that this class operator become
$G^{(1)}$ in the limit that $\Delta$ becomes $\R^3$,
as expected.

As one of $x^{\pp}$ or $x'$ is moved from the opposite
to the same side of the surface, the class operator
undergoes a discontinuity. This is reflected in the
fact that it does not satisfy the constraint. The
first crossing class operator, for example, satisfies
the equation,
$$
\left( \square^{\pp} + m^2 \right)
C^f_{\Delta} (x^{\pp}, x') =  2 \  \e (\tau - {x^0}')\
\ \delta( {x^0}^{\pp} - \tau )
\partial_0 G^{(1)}  (\tau, \x^{\pp}| {x^0}', \x')
\eqno(5.17)
$$
when $\x^{\pp}$ is in $\Delta$ and zero otherwise.
Note that Eq.(5.3) and Eq.(5.17) are consistent, since
$C_{\Delta} + C_r  = G^{(1)} $ when $\Delta = \R^3$,
and $G^{(1)}$ satisfies the constraint.
As in Section 5(A),
some doctoring of this basic class operator must therefore
be carried out
before we get the final expression, for a modified
class operator
$C_{\Delta}'$,
which satisfies the constraint.
It is, however, clear in this case how to proceed.
From the above that the obvious candidate is
to take $C_{\Delta}'$ to be given by Eq.(5.16) for {\it all}
values of the end-points $x^{\pp}$, $x'$, whether they lie on the
same side of the surface or opposite sides.
This is clearly a solution to
the constraint everywhere. It matches the path integral-defined
object when the end-points are on opposite sides of the surface.
Furthermore, when $\Delta = \R^3$, it is equal to $G^{(1)}$,
so is consistent with the modified class operator for not
crossing. We wil return below to the question of a more general
prescription for constructing modified class operators.

We may now consider the decoherence functional for histories
which cross the spacelike surface $x^0 = \tau $ either
in the region $\Delta$ or in its complement $\bar \Delta$,
for an initial state $\psi = \psi_+ + \psi_-$.
The off-diagonal terms of the decoherence functional are
$$
D(\Delta, \bar \Delta )
= \psi^* \circ_I (C_{\bar \Delta}')^{\dag} \circ_I C_{\Delta}' \circ_I \psi
\eqno(5.18)
$$
We have
$$
C_{\Delta}' \circ_I \psi
= i \int_{\Delta} d^3 x
\ \left( G^+ (x^{\pp},x) \ulra{\partial_0} \psi_+ (x)
+ G^-(x^{\pp}, x) \ulra{\partial_0} \psi_- (x) \right)
\eqno(5.19)
$$
so the decoherence functional is
$$
D( \Delta, \bar \Delta )
= \int_{\Delta} d^3 x \int_{\bar \Delta} d^3 y
\left( \psi^*_+ (y) \ulra{\partial_0} G^+(y,x)
\ulra{\partial_0} \psi_+ (x)
+ \psi^*_- (y) \ulra{\partial_0} G^-(y,x)
\ulra{\partial_0} \psi_- (x)
\right)
\eqno(5.20)
$$
The key feature of this expression is that the
decoherence functional is {\it not} exactly diagonal.
It is, however, approximately diagonal in the sense
that the two-point functions $G^{\pm} (y,x)$ decay
for increasing spatial separations. In particular,
we expect that approximate diagonality can be obtained
if both regions $\Delta $ and $\bar \Delta$ are much
larger than the Compton wavelength $m^{-1}$
(the decay length scale of $G^{\pm}(x,y)$).

Given decoherence, the probability for crossing
$\Delta$  is then
$$
\psi^* \circ_I C_{\Delta}'  \circ_I \psi
= i \int_{\Delta} d^3 x
\ \left( \psi^*_+ (x) \ulra{\partial_0} \psi_+ (x)
- \psi^*_- (x) \ulra{\partial_0} \psi_- (x) \right)
\eqno(5.21)
$$
This is exactly the expected answer, coinciding with Eq.(4.21),
although recall that
the probabilities are only approximately defined because
of approximate decoherence.

\subhead{\bf 5(D). A General Prescription for Constructing
Modified Class Operators}

We have so far constructed the modified class operators
using some general arguments, but the question remains
as to whether it is possible to find a more general
formula for constructing them. Connected to this is the question
of how the modified class operators are related to the original
path operators, such as Eq.(5.10), which were defined using
path integrals in a simple and obvious way.

Consider first, therefore, the question of why the expressions
(5.10) and (5.11) fail to satisfy the constraint. Using
the fact that the propagators of the form $g(x,t|x',0)$
satisfy the Schr\"odinger equation, it is easy to see that
(5.10) fails to satisfy the constraint at $x^{\pp}$ because
of the finite integration range for $t_c$. Recall that
$t_c$ is the parameter time of first crossing and is not
a physically observable quantity. Because it is unobservable,
and because the total parameter time $T$ is integrated
over an infinite range, it seems reasonable to explore
the possibility that $t_c$ could also be integrated over an
infinite range, in such a way that a solution to the constraint
equation is obtained.

Proceeding along these lines, one can see
that one way to obtain a solution to the constraint
is to extend the integration range of $t_c$ to
$ - \infty < t_c < \infty $, but with a signum function
$\e (t_c)$ included. That is, we define the modified class
operator
$$
C_{\Delta}' (x^{\pp}, x') = 2i \ \e (\tau - {x^0}')
\ \int_{-\infty}^{\infty} d T
\int_{-\infty}^{\infty} dt_c \ \e (t_c)
\ \int_{\Delta} d^3 x  \ g(x^{\pp},T | x, t_c)
\  \ura{\partial_0} g (x, t_c | x', 0 )
\eqno(5.22)
$$
Now we note that
$$
\eqalignno{
\int_{-\infty}^{\infty} dt_c \ \e(t_c) \ \ g (x, t_c | x', 0 )
&=
i G_F (x,x') + i G_F^* (x,x')
\cr
&= i \e (\tau - {x^0}') \ G (x,x')
&(5.23) \cr}
$$
(where recall $x^0 = \tau $).
Furthermore, we have that
$$
\ura{\partial_0} \left( \e (\tau - {x^0}') \ G (x,x') \right)
= \e (\tau - {x^0}') \ \ura{\partial_0}G (x,x')
\eqno(5.24)
$$
since $G(x,x')$ vanishes when $  \tau = {x^0}'$.
The modified class operator is therefore
$$
C_{\Delta}' (x^{\pp}, x') = -2 \int_{\Delta} d^3 x
\ G^{(1)} (x^{\pp},x)  \ura{\partial_0}G (x,x')
\eqno(5.25)
$$
which, apart from the factor of $2$, coincides with the first
term in Eq.(5.13). It is easy to see that the second term
in Eq.(5.13), but without the $\e$ factors (and again up to
a factor of $2$)
may be obtained by a slightly different modification
of the integration range over parameter times, that is,
$$
\eqalignno{
2i \ \e (\tau - {x^0}')
\ \int_{-\infty}^{\infty} d T \ \e( T- t_c)
& \ \int_{-\infty}^{\infty} dt_c \
\ \int_{\Delta} d^3 x  \ g(x^{\pp},T | x, t_c)
\  \ura{\partial_0} g (x, t_c | x', 0 )
\cr
& = -2 \int_{\Delta} d^3 x
\ G (x^{\pp},x)  \ura{\partial_0}G^{(1)} (x,x')
&(5.26) \cr}
$$
Hence averaging these two results gives the class operator
Eq.(5.13), but crucially, without the $\e$ factors that cause
(5.13) to fail to satisfy the constraint. We may finally perform
a further averaging with the last crossing versions of the above
modified class operators to obtained a modified class
operator with the following properties: it satisfies the constraint
everywhere, and, when $x^{\pp}$ and $x'$ are on
opposite sides of the surface,
coincides with the path-integral defined object
(5.16).

In summary, we therefore now have two different methods of defining
modified class operators which satisfy the constraint. One is to
use the usual path integral construction to compute the
class operators when $x^{\pp}$ and $x'$ are on opposite sides
of the surface, and then declare that this is valid for all
values of $x^{\pp}$, $x'$. The second is to use the path integral
defined object but modify the integrations over the unphysical
parameter time labels, as in Eqs.(5.22), (5.26). The two methods
are equivalent for the simple model of this paper. The benefit
of introducing the second method is that it shows that the
modification procedure does not fundamentally modify the
class of paths in configuration space summed over in the
path integral, only the way their parametrizations
are summed over.
The modified class operators therefore still have equal
claim to be a sum over paths which cross the surface.

Theses issues concerning modified class operators will be
taken up in more detail in Ref.[\cite{HaTh}].

\subhead{\bf 5(E). A Multiple Crossings Decomposition?}

The results of Section 5(C) show that it is a partition of paths
according to their first and last crossing of a spacelike
surface that leads to the expected Klein-Gordon probability
expression Eq.(4.21). It is, however, of interest to explore
other notions of surface crossings. This is partly by way of a
digression, but it is also relevant to the recovery of the
Newton-Wigner probability expression.

We begin by considering the first-crossing
path decomposition expansion Eq.(5.6),
with $T$ integrated over a half-infinite range.
We restrict to the case ${x^0}^{\pp} > \tau > {x^0}' $, and for
simplicity restrict to the positive frequency sector.
In this subsection we will take $\Delta = \R^3$.
We thus obtain
$$
G^+(x^{\pp},x') = 2 i\int d^3 x
\ G^+ (x^{\pp}, x) \ura{\partial_0} G^+ (x,x')
\eqno(5.27)
$$
(averaged with the last crossing PDX this gives the composition law
in Eq.(3.2)).
From Section 3 we also have that
$$
2i \partial_0 G^+ (x^{\pp},x')
= G_{NW} (x^{\pp},x')
\eqno(5.28)
$$
and Eq.(5.27) becomes
$$
G^+ (x^{\pp},x') =  \int d^3 x
\ G^+ (x^{\pp}, x) G_{NW} (x,x')
\eqno(5.29)
$$
In this expression the Newton-Wigner propagator therefore
represents paths that start at $x'$, move forwards and backwards
in time but without crossing the final surface,
except to end on it at point $x$. (See Fig.(3.2)).
Similarly, one can see
from the last crossing PDX Eq.(5.7), that a sum over
paths which move backwards and forwards in time
but without crossing the initial surface, except to start
on it, is $ - 2 i G^+ (x^{\pp},x') \ula{\partial_0} $.
(See Fig.(5.4)).
This is in fact again the Newton-Wigner propagator $G_{NW}(x,x')$,
as may be seen from Eq.(3.13),
and we may write
$$
G^+ (x^{\pp},x') =  \int d^3 x
\ G_{NW} (x^{\pp}, x) G^+ (x,x')
\eqno(5.30)
$$
Using these two relations, we may carry out an iteration of Eq.(5.29)
to yield
$$
G^+ (x^{\pp}, x')
=\int d^3 x_1 \ d^3 x_2 \ G_{NW} (x^{\pp},x_2) G^+ (x_2, x_1)
G_{NW} (x_1,x')
\eqno(5.31)
$$
The two Newton-Wigner propagators represent restricted
propagation to the first crossing and from the last
crossing of the surface, and the propagator $G^+ (x_2,x_1)$
is unrestricted propagation between two points which
both lie on the spacelike surface. (See Fig.(5.5)).

It is now reasonable to ask whether this expression can be
further decomposed according to the detailed number of surface crossings
entailed in the path integral representation of
$G^+ (x_2, x_1)$. Interestingly, this does not appear to be
possible. For suppose we apply a first or last crossing
expansion of the type Eq.(5.29) or Eq.(5.30) to the propagator
$G^+ (x_2, x_1)$. The point is that all three
points involved (initial point, final point, crossing point)
all have the same value of $x^0$, hence the expression would
contain the Newton-Wigner propagator $G_{NW} (x,y)$
at $x^0 = y^0$. But this is simply the delta
function $\delta^{(3)} (\x - \y)$, leading to a trivial result.
Hence a further decomposition according to the specific number
of crossings appears to be impossible.

The explanation for this is that in a path integral representation
of $G^+(x_2, x_1)$, generic paths cross the surface an infinite
number of times, and the set of paths crossing a finite number of
times is of measure zero. This result is due in essence to Hartle
[\cite{Har1}], who considered a lattice version of the Euclideanized sum
over histories. He attempted to factor the usual propagator of
non-relativistic quantum mechanics across an arbitrary surface in
spacetime by partitioning according to the number of crossings
each path makes. He showed that paths with a finite number of
crossings generally have zero amplitude in the continuum limit,
and deduced that such a factoring is not in fact possible. (This
is not at variance with the path decomposition expansion,
Eq.(5.6), which partitions the paths according to their first
crossing).

On the face of it, therefore, it might seem like there are a
number of different notions of surface crossing. The above results show,
however, that first and last crossings are the only ones that can be
defined in this case, hence are the only useful ones for defining
the class operators of interest here.

\head{\bf 6. The Newton-Wigner Case}

Consider now the question of how to obtain the Newton-Wigner
probability Eq.(4.13) from the decoherent histories approach.
Since this is like non-relativistic quantum mechanics but
with the Hamiltonian $ h = \sqrt{ - \nabla^2 +m^2 }$,
it is simpler than the previous case and we describe it only
briefly.

The decoherence functional is given by Eq.(3.17) in which
the inner product $\circ$ is the Schr\"odinger inner product,
with the initial states normalized in this product. The main issue
is the construction of the class operator representing
crossing a spacelike surface $x^0 = \tau$ in a spatial region $\Delta$.
We first give the result and then explain its origin. (For
simplicity we concentrate on the positive frequency sector only).
It is clear that the class operator is
$$
C_{\Delta} (x^{\pp}, x') = \int_{\Delta} d^3 x \ G_{NW} (x^{\pp},\x, \tau)
\ G_{NW} (\x, \tau ,x')
\eqno(6.1)
$$
Note that when $\Delta = \R^3 $ this gives the standard
composition property of the Newton-Wigner propagator
$$
G_{NW} (x^{\pp},x') = \int d^3 x \ G_{NW} (x^{\pp},\x, \tau ) \
G_{NW} (\x, \tau ,x')
\eqno(6.2)
$$
Inserting in the decoherence functional (3.17), it is
readily shown this gives exact
decoherence, and the probabilities coincide with Eq.(4.13)
(this is very similar to standard calculations in non-relativistic
quantum mechanics). It is necessary also to use here
the fact that the Newton-Wigner propagator is the overlap of two
Newton-Wigner states,
$$
G_{NW} (\x, x^0 | \y, y^0 ) = \la \x, x^0 | \y, y^0 \ra
\eqno(6.3)
$$

There are (at least) two path integral representations of the path
integral for the Newton-Wigner propagator that lead to the
class operator (6.1). The first is the one of the standard non-relativistic
form:
$$
G (\x^{\pp}, \tau^{\pp},\x', \tau') = \int {\cal D} \x {\cal D} \p
\exp \left( i \int_{\tau'}^{\tau^{\pp}} dx^0 \left[ \p \cdot { d \x \over d x^0}
- \sqrt{ \p^2 +m^2 } \right] \right)
\eqno(6.4)
$$
(The configuration space form of this path integral may also be
considered, but the measure is then rather complicated [\cite{HaKu}]).
In this representation, the paths move forwards in the time coordinate
$x^0$.
Summing over paths from $x'$ to $x^{\pp}$ which pass through
$\Delta$ on an intermediate spacelike surface then yields
the class operator (6.1).

A second and perhaps more interesting path integral representation
of $G_{NW} (x^{\pp}, x')$ is the one mentioned in the discussion of the path
decomposition expansion of the previous section. This is to use a
path integral representation of the form (1.15), (1.16), in which
the paths summed over do not cross the final surface except to end
on it at $x^{\pp}$, depicted in Fig.(3.2).
(See also Refs.[\cite{Har1,HaOr}]). Equivalently, they can
be restricted so that they start at the initial point $x'$ but do
not cross it thereafter (see Fig.(5.4)).
In fact, it is easy to show from these
representations, using Eq.(6.2), that $G_{NW}(x^{\pp},x')$ is
obtained more generally by choosing {\it any} surface of constant
$x^0$ lying between initial and final points, and then summing
over paths which cross it once and only once, as depicted
in Fig.(5.6). The first
two representations then correspond to the limit in which the
intermediate surface tends to the initial or final surface.
From this third representation, we see that the class operator
Eq.(6.1) is obtained by summing over paths which cross the
intermediate spacelike surface once and only once, in the spatial
region $\Delta$.

Note that there is no conflict here with the statement in Section
5(D) that the set of paths crossing a surface only a finite number
of times is of zero measure in the set of all paths. Section 5(D)
concerned path integral representations of the Klein-Gordon
propagators, which involve a sum over {\it all} paths between two
points, and the set of paths making single crossings of given
surface are indeed insignificant. However, the path integral
representation of the Newton-Wigner propagator considered is
defined from the outset by a sum over the much smaller class
of paths which cross an intermediate surface only once.

\head{\bf 7. Summary and Discussion}

This paper is a first step in a programme whose general aim
is to supply a reasonable predictive framework for quantum
cosmological models. In connection with that aim, the main
achievement of this paper is the derivation of the
probability formula Eq.(5.21) (or Eq.(4.21)) from the
decoherent histories approach, and the demonstration
that the associated histories are approximately decoherent.

Along the way we derived a number of other relevant results.
We showed how to modify in a physically sensible way
class operators which do not satisfy the constraints.
We also showed that first and last crossings are
essentially the only ways of defining surface crossings
(in the Klein-Gordon quantization). In particular,
partitions of paths according to multiple crossings
are not possible. These results will be relevant
to more complicated models in quantum cosmology.

The main result of Section 4 is the computation of an evolving
constraints operator for the relativistic particle, proof that its
eigenstates are the Newton-Wigner states, and that the operator is
essentially the same as the Newton-Wigner operator. We discussed
some novel path integral representations of the NW propagator in
Section 6, involving single surface crossings and computed the
decoherence functional (although noted that it is very similar to
the case of non-relativistic quantum mechanics).

We do not expect that a Newton-Wigner quantization based on a
Schr\"odinger equation such as Eq.(3.12) will be relevant to more
complicated models in quantum cosmology, since it is only under
very special circumstances that the constraint may be solved to
produce a real, positive Hamiltonian $h$ to go in Eq.(3.12). The
comparison with this case has, however, proved quite useful in the
present paper. Furthermore, here, our starting point for an
operator quantization was the evolving constants method, based on
Eq.(1.8), which does not require the solution to the constraints,
and this method will be valid for more complicated models (and
indeed has been used already in such a context [\cite{Mar1,Mar2}]).
We also note that we find agreement between the operator
methods of Section 4 and the decoherent histories results
of Sections 5 and 6.

We may now return to the discussion initiated at the end of
Section 2, on the relationship between decoherence, surface
crossings and the existence of self-adjoint operators.
We see further evidence for this.
It is striking that the NW quantization, which
involves single surface crossings in the decoherent histories
approach, yields exact decoherence of histories, whereas the KG
quantization, which has multiple surface crossings, exhibits only
approximate decoherence. It is reasonable to conclude from this
that the approximate nature of the decoherence is related to
the fact that the paths in KG quantization go backwards and
fowards in time and cross a surface many times. In fact, generally
speaking, one might expect such multiple crossings to destroy
decoherence altogether, since the paths may pass through both
$\Delta$ and its complement $\bar \Delta$, but with single
crossings they may pass through only one or the other. The
interesting question is therefore why even approximate decoherence
is obtained. For the free relativistic particle considered here,
the answer is that the dominant contribution to the path integral
representation of the class operators comes from the immediate
neighbourhood of the {\it classical} path from $x'$ to $x^{\pp}$,
and this crosses the surface only once. Paths with multiple
crossings therefore presumably belong to the quantum fluctuations
about the classical path, and these may be neglected at
sufficiently coarse-grained scales.
Note also that the anticipated connection with self-adjoint
operators holds up. The NW probability is associated with
a self-adjoint operator, whilst the KG probability is not.

These issues concerning multiple crossings become more complicated
in non-trivial quantum cosmological models where typically even
the classical paths have multiple surface crossings. This will be
pursued elsewhere [\cite{HaTh}].

\head{\bf Acknowledgements}

We are very grateful to Jim Hartle and Don Marolf
for useful conversations. JT was supported by the
Evangelisches Studienwerk Villigst.

\references

\def\pr{{\sl Phys. Rev.\ }}

\def\jmp{{\sl J. Math. Phys.\ }}

\def\np{{\sl Nucl. Phys.\ }}

\def\annp{{\sl Ann. Phys. (N.Y.)\ }}
\def\cqg{{\sl Class. Quant. Grav.\ }}

\refis{AnSa} C.Anastopoulos and K.Savvidou, in preparation.

\refis{Ash} For a nice review see, C.Rovelli, gr-qc/9710008,
{\it Loop quantum gravity}.

\refis{BuIs} J.Butterfield and C.J.Isham, gr-qc/9901024.
%{\it On the emergence of time in quantum gravity}.

\refis{CrHa} D.Craig and J.B.Hartle, unpublished.
preprint UCSBTH-94-47 (1998).
% Generalized quantum theory of Bianchi IX Cosmologies.

\refis{DeW} B.DeWitt, in {\it Gravitation: An Introduction to
Current Research}, edited by L.Witten (John WIley and Sons, New
York, 1962).
% The quantization of geometry.

\refis{GeH} M.Gell-Mann and J.B.Hartle, in {\it Complexity,
Entropy and the Physics of Information, SFI Studies in the
Sciences of Complexity}, Vol. VIII, W. Zurek (ed.) (Addison
Wesley, Reading, 1990); and in {\it Proceedings of the Third
International Symposium on the Foundations of Quantum Mechanics in
the Light of New Technology}, S. Kobayashi, H. Ezawa, Y. Murayama
and S. Nomura (eds.) (Physical Society of Japan, Tokyo, 1990);
%{\it Quantum Mechanics in the Light of Quantum Cosmology.}
{\sl Phys.Rev.} {\bf D47}, 3345 (1993).

\refis{Gri} R.B.Griffiths, {\sl J.Stat.Phys.} {\bf 36}, 219
(1984); {\sl Phys.Rev.Lett.} {\bf 70}, 2201 (1993); {\sl
Am.J.Phys.} {\bf 55}, 11 (1987).

\refis{Omn} R.Omn\`es, {\sl J.Stat.Phys.} {\bf 53}, 893 (1988);
{\bf 53}, 933 (1988); {\bf 53}, 957 (1988); {\bf 57}, 357 (1989);
{\bf 62}, 841 (1991); {\sl Ann.Phys.} {\bf 201}, 354 (1990); {\sl
Rev.Mod.Phys.} {\bf 64}, 339 (1992).

\refis{IsL} For further developments in the decoherent histories
approach, particularly adpated to the problem of spacetime coarse
grainings, see C. Isham, \jmp {\bf 23}, 2157 (1994);
%{\it Quantum Logic and the Histories Approach to Quantum Theory.}
C. Isham and N. Linden, \jmp {\bf 35}, 5452 (1994); {\bf 36}, 5392
(1995).

\refis{GRT} N.Grot, C.Rovelli and R.S.Tate, \pr {\bf A54}, 46
(1996).

\refis{Hal1} J.J.Halliwell, in, {\it Proceedings of the 13th
International Conference on General Relativity and Gravitation},
edited by R.J.Gleiser, C.N.Kozameh, O.M.Moreschi
(IOP Publishers, Bristol,1992). (Also available as
the e-print gr-qc/9208001).
% The Interpretation of Quantum Cosmological Models

\refis{Hal2} Yet another approach to interpreting
the Wheeler-DeWitt equation is to use a model detector
in the Hamiltonian. See J.J.Halliwell, e-print gr-qc/0008046
(accepted for publiation in Physical Review D).

\refis{Hal3} J.J.Halliwell, \pr {\bf D38}, 2468 (1988).
% Derivation of the Wheeler-DeWitt equation
% from a Path Integral for Minisuperspace Models.

%\refis{Hal4} This detector model was used in a simple
%non-relativistic context by J.J.Halliwell, \pr {\bf D60}, 105031
%(1999).
% ``Somewhere in the Universe: Where is the Information Stored when
% Histories Decohere?'', quant-ph/9902008, Imperial preprint
% TP/98-99/29 (1999).
%Some subsequent developments of the Coleman-Hepp model are
%H.Nakazato and S.Pascazio, \prl {\bf 70}, 1 (1993); \pr {\bf A48},
%1066 (1993); R.Blasi, S.Pascazio, S.Takagi, \pr {\bf A250}, 230 (1998).

%\refis{Hal5} J.J.Halliwell, {\sl Prog.Theor.Phys.} {\bf 102}, 707 (1999).

\refis{Hal6} J.J.Halliwell, {\sl Phys.Rev.} {\bf D60}, 105031 (1999).
%``Somewhere in the Universe: Where is the Information
%Stored when Histories Decohere?'',
%Imperial/TP/98--99/29. quant-ph/9902008.

\refis{HalH} J.J.Halliwell and J.B.Hartle, {\sl Phys.Rev.}
{\bf D43}, 1170 (1991).
% ``Wave Functions Constructed from an Invariant
% Sum-Over-Histories Satisfy Constraints",

\refis{HaOr} J.J.Halliwell and M.E.Ortiz, {\sl Phys.Rev.} {\bf D48}, 748 (1993).

\refis{HaTh} J.J.Halliwell and J.Thorwart, in preparation.

\refis{Whe} J.Whelan, \pr {\bf D50}, 6344 (1994);
% Spacetime alternatives in relativistic particle motion

\refis{HaMa} J.B.Hartle and D.Marolf,  \pr {\bf D56}, 6247 (1997).
% Comparing Formulations of Generalized Quantum Mechanics for
% Reparametrization-Invariant Systems

\refis{HaKu} J.B.Hartle and K.Kuchar, \pr {\bf D34}, 2323 (1986).
% Path integrals in parametrized theories: the free relativistic
% particle

\refis{Har1} J.B.Hartle, \pr {\bf D37}, 2818 (1988).
% Quantum kinematics of spacetime. I. Non-relativistic theory.

%\refis{Har2} J.B.Hartle, \pr {\bf D38}, 2985 (1988).
% Quantum kinematics of spacetime. II. A model quantum cosmology
% with real clocks.

\refis{Har3} J.B.Hartle, in {\it Proceedings of the 1992 Les Houches
School, Gravity and its Quantizations}, edited by B.Julia and
J.Zinn-Justin (Elsevier Science B.V. 1995), also
available as the e-print, gr-qc/9304006.
% Spacetime quantum mechanics and the quantum mechanics of
% spacetime

\refis{HaHa} J.B.Hartle and S.W.Hawking, \pr {\bf D28}, 2960
(1983).

\refis{Ish} C.J.Isham, gr-qc/9210011.
%{\it Canonical quantum gravity and the problem of time}.

\refis{Kuc} K.Kuchar, in {\it Conceptual Problems of Quantum
Gravity}, edited by A.Ashtekar and J.Stachel (Boston, Birkhauser,
1991); and in {\it Proceedings of the 4th Canadian Conference on
General Relativty and Relativistic Astrophysics}, edited by
G.Kunstatter, D.E.Vincent and J.G.Williams (World Scientific, New
Jersey, 1992). See also the e-print gr-qc/9304012,
{\it Canonical quantum gravity}.

\refis{Mar1} D.Marolf, \cqg {\bf 12}, 1199 (1995).
% Quantum observables and recollapsing dynamics

\refis{Mar2} D.Marolf, \pr {\bf D53}, 6979(1996);
% Path Integrals and Instantons in Quantum Gravity
\cqg {\bf 12}, 2469 (1995);
% Almost Ideal Clocks in Quantum Cosmology: A Brief Derivation of Time
\cqg{\bf 12}, 1441 (1995).
% Observables and a Hilbert Space for Bianchi IX

\refis{NeW} T.D.Newton and E.P.Wigner, \pr {\bf 21}, 400 (1949).
% Localized states for elementary systems

\refis{PDX} A.Auerbach and S.Kivelson, \np {\bf B257}, 799 (1985).
%The path decomposition expansion and multidimensional tunneling

\refis{Rie} A.Ashtekar, J.Lewandowski, D.Marolf, J.Mourao and
T.Thiemann, {\sl J.Math.Phys.} {\bf 36}, 6456 (1995);
% Quantization of diffeomorphism invariant theorie of connections with
% local degrees of freedom.
A.Higuchi, \cqg {\bf 8}, 1983 (1991).
% Quantum linearization instabilities of de Sitter spacetime 2.
D.Giulini and D.Marolf, \cqg {\bf 16}, 2489 (1999);
% A Uniqueness Theorem for Constraint Quantization
\cqg {\bf 16}, 2479 (1999).
% On the Generality of Refined Algebraic Quantization
F.Embacher, {\sl Hadronic J.} {\bf 21}, 337 (1998);
% Handwaving refined algebraic quantization.
N.Landsmann, {\sl J.Geom.Phys.} {\bf 15}, 285 (1995).
% Rieffel induction as generalized quantum Marsden-Weinstein
% reduction

\refis{Rov1} C.Rovelli, \pr {\bf D42}, 2638 (1990).
% Quantum mechanics without time: A model.

\refis{Rov2} C.Rovelli, \pr {\bf D43}, 442 (1991).
% Time in quantum gravity: An hypothesis.

\refis{Rov3} C.Rovelli, \cqg {\bf 8}, 297 (1991);
% What is observable in classical and quantum gravity
{\bf 8}, 317 (1991).
% Quantum references systems.

\refis{Time}
Y.Aharanov and D.Bohm, \pr {\bf 122}, 1649 (1961);
% Time in quantum theory and the uncertainty relation for time
Y.Aharanov, J.Oppenheim, S.Popescu, B.Reznik and
W.Unruh, quant-ph/9709031 (1997);
% Measurement of Time-of-Arrival in Quantum Mechanics.
G.R.Allcock, \annp {\bf 53}, 253 (1969);
{\bf 53}, 286 (1969); {\bf 53}, 311 (1969);
% The Time of Arrival in Quantum Mechanics
Ph.Blanchard and A.Jadczyk,
{\sl Helv.Phys.Acta.} {\bf 69}, 613 (1996);
% Time of events in quantum theory.
I.Bloch and D.A.Burba, \pr {\bf 10}, 3206 (1974);
% Presence of a particle in a given sacpe-time region and the
% continuous action of a particle dectector.
V.Delgado, preprint quant-ph/9709037 (1997);
R.Giannitrapani, preprint quant-ph/9611015 (1998);
%N.Grot, C.Rovelli and R.S.Tate, \pr {\bf A54}, 46 (1996);
E.Gurjoy and D.Coon, {\sl Superlattices and
Microsctructures} {\bf 5}, 305 (1989);
A.S.Holevo, {\it Probabilistic and Statistical Aspects
of Quantum Theory} (North Holland, Amsterdam, 1982), pages 130--197;
A.Jadcyk, {\sl Prog.Theor.Phys.} {\bf 93}, 631 (1995);
% Particle Tracks, Events and Quantum Theory.
D.H.Kobe and V.C.Aguilera--Navarro,  \pr {\bf A50}, 933 (1994);
N.Kumar, {\sl Pramana J.Phys.} {\bf 25}, 363 (1985);
J.Le\'on, preprint quant-ph/9608013 (1996);
D.Marolf, \pr {\bf A50}, 939 (1994);
L.Mandelstamm and I.Tamm, {\sl J.Phys.} {\bf 9}, 249 (1945);
J.G.Muga, S.Brouard and D.Mac\'ias, \annp {\bf 240}, 351 (1995);
% Time of arrival in quantum mechanics.
J.G.Muga, J.P.Palao and C.R.Leavens, preprint quant-ph/9803087
(1987);
J.G.Muga, R.Sala and J.P.Palao, preprint quant-ph/9801043,
{\sl Superlattices and Microstructures} {\bf 23} 833 (1998);
C.Piron, in {\it
Interpretation and Foundations of Quantum Theory}, edited by
H.Newmann (Bibliographisches Institute, Mannheim, 1979);
M.Toller, preprint quant-ph/9805030 (1998);
H.Salecker and E.P.Wigner, \pr {\bf 109}, 571 (1958);
F.T.Smith, \pr {\bf 118}, 349 (1960);
E.P.Wigner, \pr {\bf 98}, 145 (1955).

\refis{YaT} N.Yamada and S.Takagi, {\sl Prog.Theor.Phys.}
{\bf 85}, 985 (1991); {\bf 86}, 599 (1991); {\bf 87}, 77 (1992);
N. Yamada, {\sl Sci. Rep. T\^ohoku Uni., Series 8}, {\bf 12}, 177
(1992); \pr {\bf A54}, 182 (1996); J.J.Halliwell and E.Zafiris,
{\sl Phys.Rev.} {\bf D57}, 3351-3364 (1998);
J.B.Hartle, {\sl Phys.Rev.} {\bf D44}, 3173 (1991);
%{\it Spacetime Coarse-Grainings in Non-Relativistic Quantum Mechanics.}
R.J.Micanek and J.B.Hartle, {\sl Phys.Rev.} {\bf A54},
3795 (1996).
% Nearly instantaneous alternatives in quantum mechanics.

\endreferences

\endpage

\head{\bf Figures}

\epsfbox{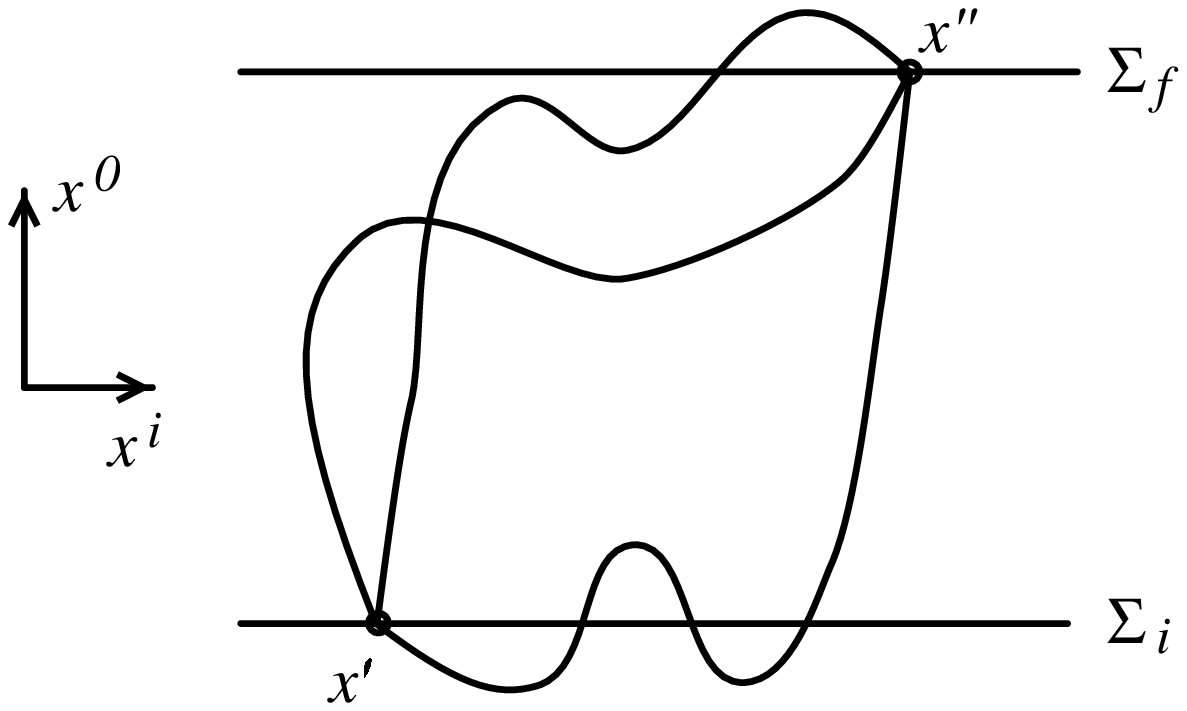}

\noindent{\bf Figure 3.1} The propagators $G^{(1)} (x^{\pp}, x')$
and $G_F (x^{\pp},x')$ are obtained by a path integral of
the form (1.15), (1.16), involving a sum over {\it all} paths
from $x'$ to $x^{\pp}$ (hence the paths may move both backwards
and forwards in the time coordinate $x^0$). An infinite range
for $T$ gives $G^{(1)} (x^{\pp}, x')$ and a half-infinite
positive range gives $ i G_F (x^{\pp} ,x')$.

\epsfbox{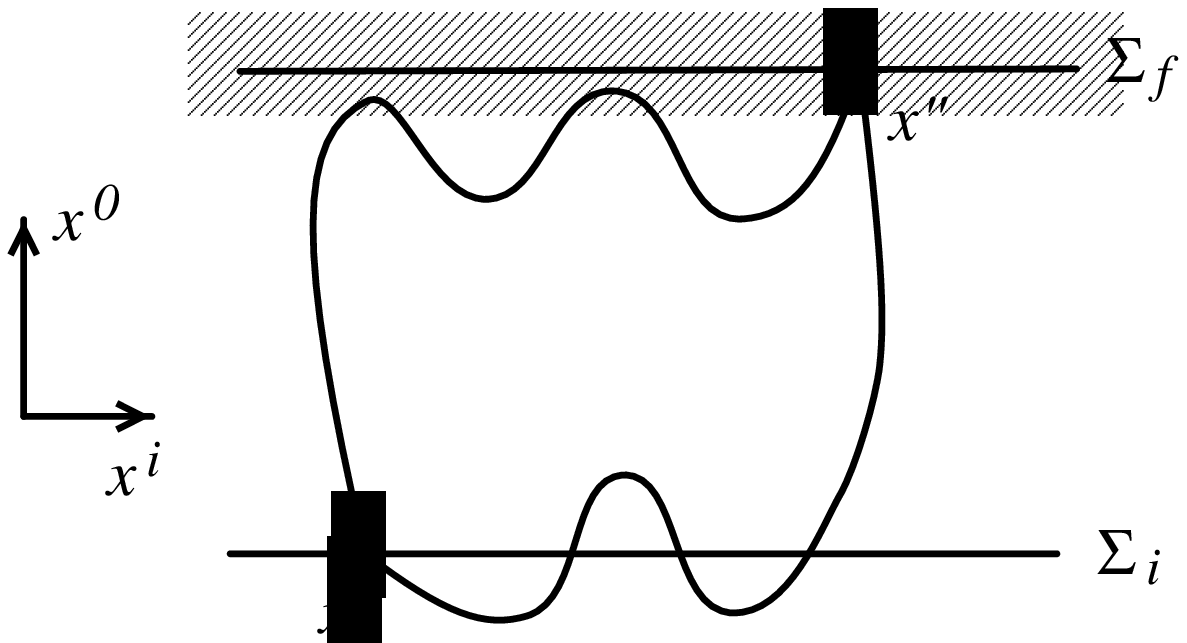}

\noindent{\bf Figure 3.2} The Newton-Wigner propagator
$G_{NW} (x^{\pp}, x')$ may be obtained by a path integral of
the form (1.15), (1.16), in which the paths may move backwards
and forwards in time, with the restriction that
they do not cross
the final surface, except to end on it at the final point $x^{\pp}$.

\endpage

\epsfbox{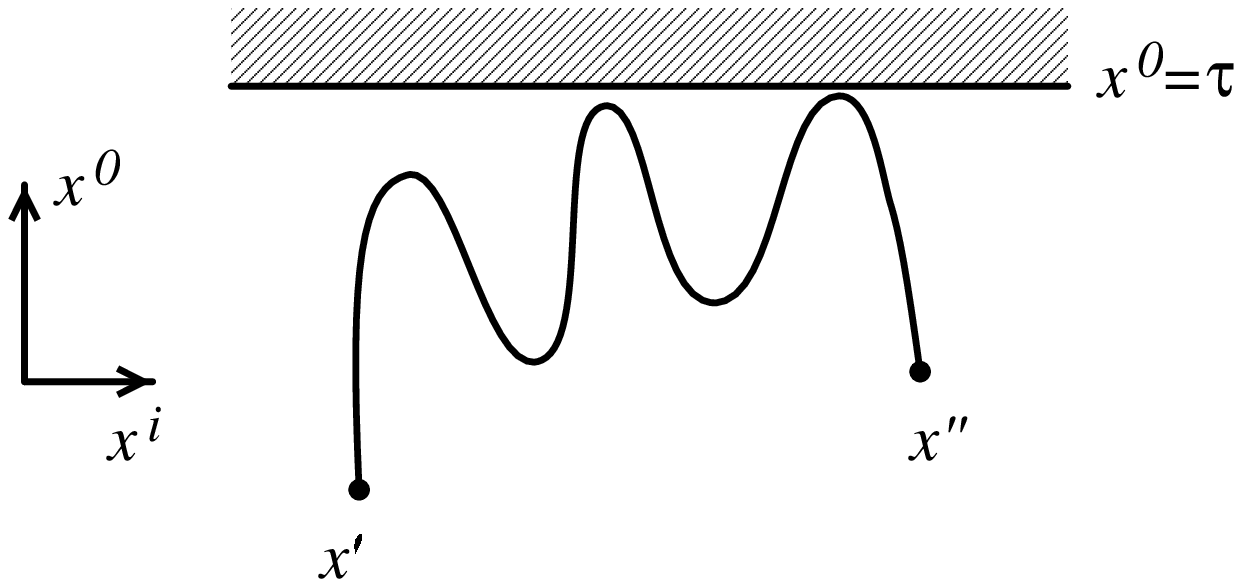}

\noindent{\bf Figure 5.1} The class operator $C_r (x^{\pp},x')$
for not crossing the spacelike surface $x^0 = \tau$ is given
by a sum over paths from $x'$ to $x^{\pp}$ which never cross
the surface. It is equivalent to a method of images
construction.

\bigskip
\epsfbox{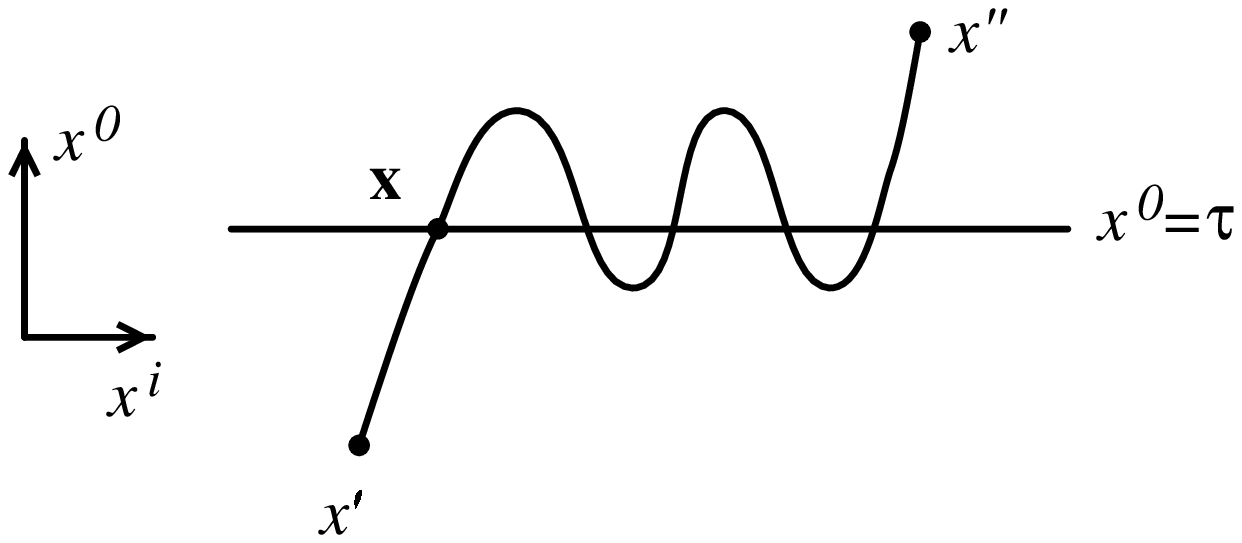}

\noindent{\bf Figure 5.2} The path decomposition expansion (PDX)
for the surface $x^0 = \tau$. A sum over paths from points $x'$ to
$x^{\pp}$ on opposite sides of a surface may be partitioned
according to the position ${\bf x}$ and the parameter time
at which it makes its {\it first} crossing of the surface.

\bigskip

\epsfbox{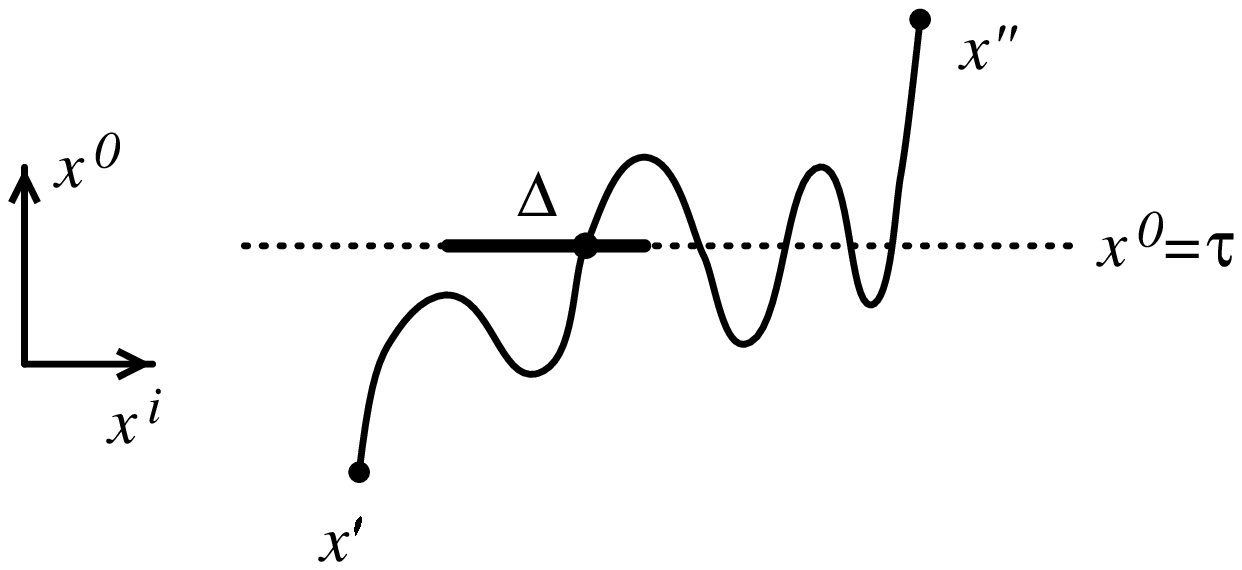}

\noindent{\bf Figure 5.3} The class operator $C_{\Delta}(x^{\pp},x')$
for a first crossing of the surface $x^0 = \tau$ in the spatial
region $\Delta$ is obtained by summing over paths which cross
the surface for the first time in the region $\Delta$.

\bigskip
\epsfbox{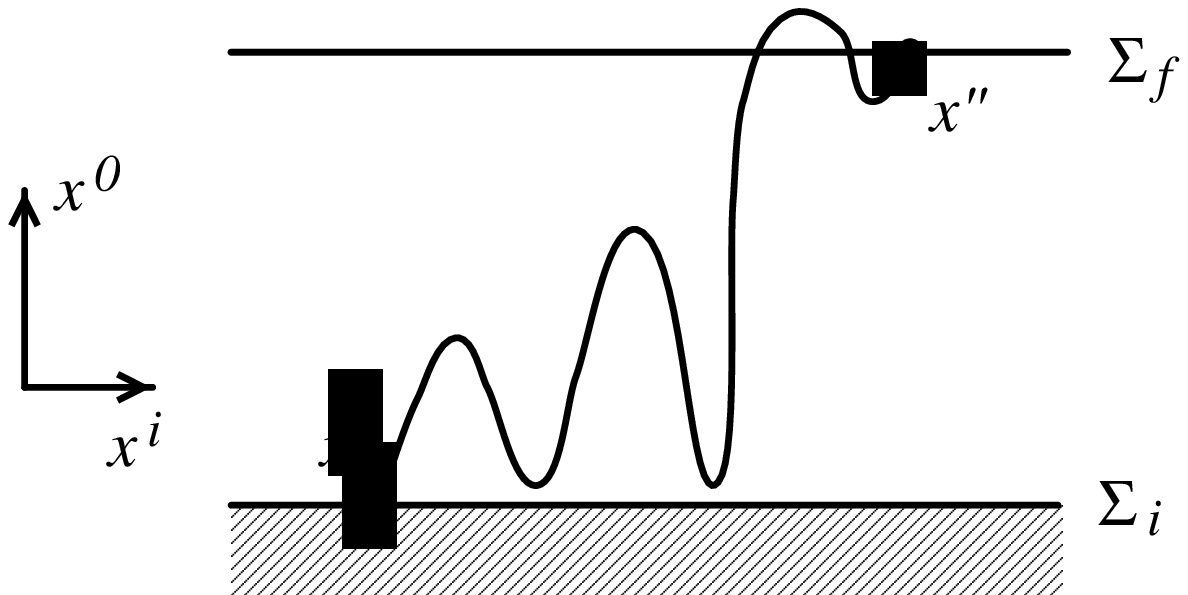}

\noindent{\bf Figure 5.4} A second representation of
the Newton-Wigner propagator
$G_{NW} (x^{\pp}, x')$ may be obtained by a path integral of
the form (1.15), (1.16), in which the paths may move backwards
and forwards in time, with the restriction that
they do not cross
the initial surface, except to start on it at the initial point $x'$.

\bigskip
\epsfbox{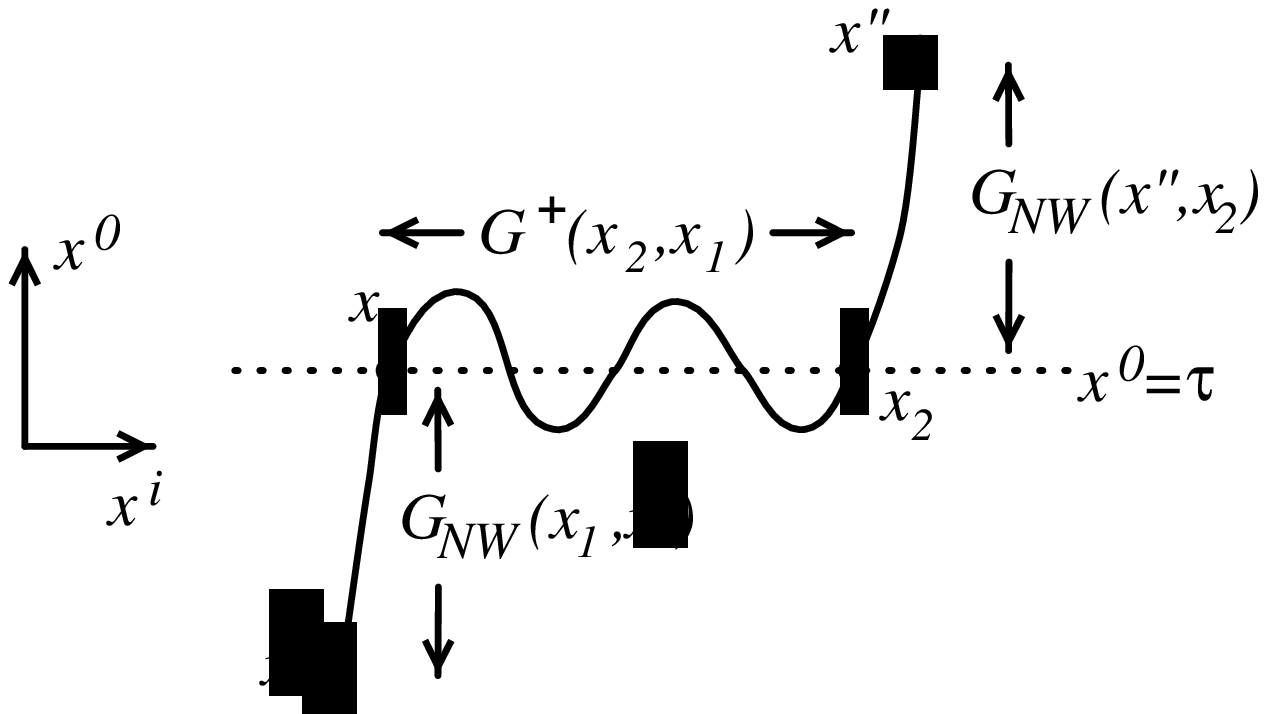}

\noindent{\bf Figure 5.5} The path integral representation of
Eq.(5.31). The paths go from the initial point $x'$ to their
first crossing of the surface at $x_1$, and this is
represented by $G_{NW}(x_1,x')$. The propagation from
$x_1$ to $x_2$ is unrestricted and is represented by
$G^+ (x_1,x_2)$. The paths make their last crossing
at $x_2$ moving to their final point $x^{\pp}$, and
this is represented by $G_{NW} (x^{\pp},x_2)$.

\bigskip
\epsfbox{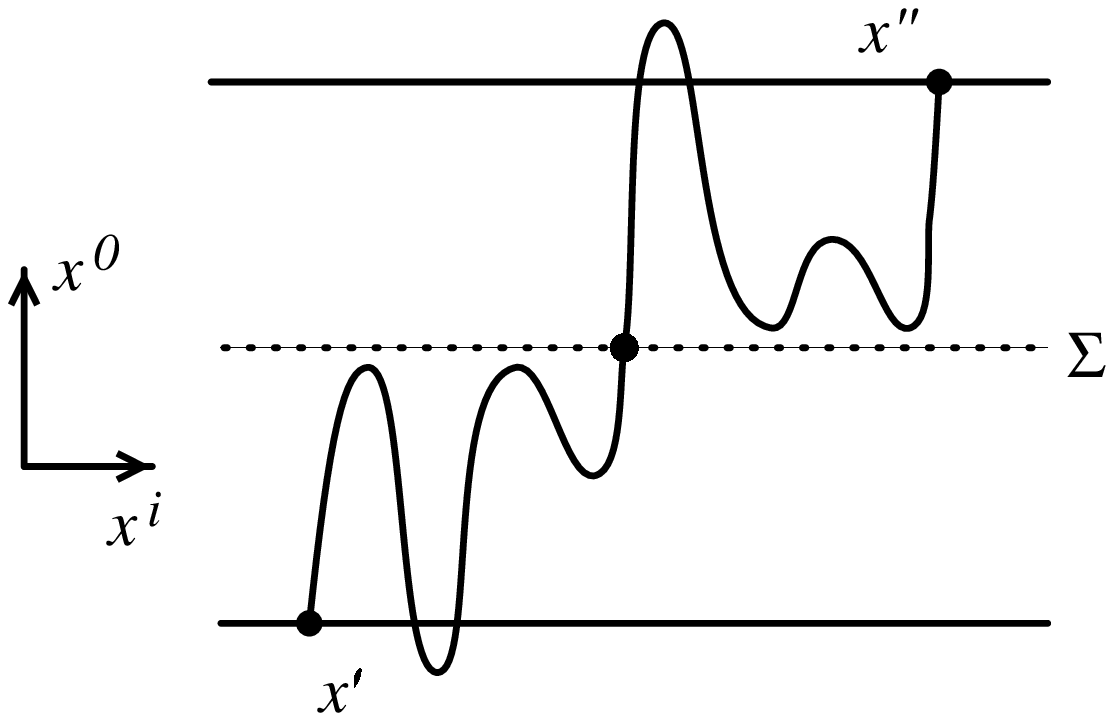}

\noindent{\bf Figure 5.6} A more general path integral
representation of the
Newton-Wigner propagator $G_{NW}(x^{\pp},x')$.
In the path integral expressions
(1.15), (1.16), the paths move backwards and forwards in
time with the restriction that they may cross a prescribed
intermediate surface once and only once. The representation
is independent of the choice of intermediate surface. The
two previous representations (portrayed in Fig.(3.2) and
Fig.(5.4)) in the limit that the intermediate surface
tends to the final or initial surface.

\end